\documentclass[useAMS,usenatbib]{mnras}
\usepackage{amsmath}
\usepackage{appendix}
\usepackage{graphicx}
\usepackage{caption}
\usepackage{array}
\usepackage[figuresright]{rotating}
\usepackage{booktabs}
\usepackage[flushleft]{threeparttable}
\usepackage[export]{adjustbox}

\allowdisplaybreaks

\newcommand{\Lagr}{\mathcal{L}}
\newcommand{\NumberGal}{21}
\newcommand{\FinalGammaValue}{$-2.12\,\pm\,0.05$}

\usepackage{etoolbox}
\makeatletter
\patchcmd\@combinedblfloats{\box\@outputbox}{\unvbox\@outputbox}{}{%
	\errmessage{\noexpand\@combinedblfloats could not be patched}%
}%
\makeatother

\title[SLUGGS: dynamical modelling to $\sim$4 effective radii]{The SLUGGS Survey: A comparison of total-mass profiles of early-type galaxies from observations and cosmological simulations, to $\sim$4 effective radii }
\author[Bellstedt et al.]
{Sabine Bellstedt,$^{1,2}$\thanks{Email: sbellstedt@swin.edu.au} 
	Duncan A. Forbes,$^{1}$ 
	Aaron J. Romanowsky,$^{3,4}$ 
	\newauthor Rhea-Silvia Remus,$^{5, 6}$  
	Adam R. H. Stevens,$^{7}$ 
	Jean P. Brodie,$^{3}$ 
	Adriano Poci,$^{8}$ 
	\newauthor Richard McDermid,$^{2,8}$ 
	Adebusola Alabi,$^{1,3}$  
	Leonie Chevalier,$^{1}$ 
	Caitlin Adams,$^{1}$ 
	\newauthor Anna Ferr\'{e}-Mateu,$^{1}$ 
	Asher Wasserman$^{3}$ 
	and Viraj Pandya,$^{3}$ \\
$^{1}$Centre for Astrophysics and Supercomputing, Swinburne University of Technology, Hawthorn VIC 3122, Australia\\
$^{2}$Australian Astronomical Observatory, PO Box 915, Sydney NSW 1670, Australia\\
$^{3}$University of California Observatories, 1156 High Street, Santa Cruz, CA 95064, USA\\
$^{4}$Department of Physics and Astronomy, San Jos\'{e} State University, One Washington Square, San Jose, CA, 95192, USA\\
$^{5}$Universit\"ats-Sternwarte M\"unchen, Fakult\"at f\"ur Physik, LMU Munich, Scheinerstr. 1, D-81679 M\"unchen, Germany\\
$^{6}$Canadian Institute for Theoretical Astrophysics, 60 St. George Street, University of Toronto, Toronto, ON, M5S 3H8, Canada\\
$^{7}$International Centre for Radio Astronomy Research, University of Western Australia, Crawley, WA, 6009, Australia\\
$^{8}$Department of Physics and Astronomy, Macquarie University, North Ryde, NSW 2109, Australia\\
}

\begin{document}

\date{}

\pagerange{\pageref{firstpage}--\pageref{lastpage}} \pubyear{2018}

\maketitle

\label{firstpage}

\begin{abstract}

We apply the Jeans Anisotropic MGE (JAM) dynamical modelling method to SAGES Legacy Unifying Globulars and GalaxieS (SLUGGS) survey data of early-type galaxies in the stellar mass range $10^{10}<M_*/{\rm M}_{\odot}<10^{11.6}$ that cover a large radial range of $0.1-4.0$ effective radii. 
We combine SLUGGS and ATLAS$^{\rm 3D}$ datasets to model the total-mass profiles of a sample of \NumberGal$\,$ fast-rotator galaxies, utilising a hyperparameter method to combine the two independent datasets. The total-mass density profile slope values derived for these galaxies are consistent with those measured in the inner regions of galaxies by other studies. Furthermore, the total-mass density slopes ($\gamma_{\rm tot}$) appear to be universal over this broad stellar mass range, with an average value of $\gamma_{\rm tot}=\,$\FinalGammaValue$\,$, i.e. slightly steeper than isothermal. 
We compare our results to model galaxies from the \emph{Magneticum} and EAGLE cosmological hydrodynamic simulations, in order to probe the mechanisms that are responsible for varying total-mass density profile slopes. The simulated-galaxy slopes are shallower than the observed values by $\sim0.1-0.3$, indicating that the physical processes shaping the mass distributions of galaxies in cosmological simulations are still incomplete. 
For galaxies with $M_*>10^{10.7}{\rm M}_{\odot}$ in the \emph{Magneticum} simulations, we identify a significant anticorrelation between total-mass density profile slopes and the fraction of stellar mass formed ex situ (i.e. accreted), whereas this anticorrelation is weaker for lower stellar masses, implying that the measured total mass density slopes for low-mass galaxies are less likely to be determined by merger activity.

\end{abstract}


\section{Introduction}

The goal of developing a complete understanding of galaxy formation has been driving studies of galaxies for decades. 
Despite the significant advances that have been made, there are still many unknowns pertaining to the formation of early-type galaxies (ETGs; elliptical and lenticular), for which multiple different scenarios have been proposed. A property of galaxies that has been of longstanding interest in better understanding these galaxies is the distribution of mass (both baryonic and dark) within the galaxy itself. 

The presentation of a mathematical description of the distribution of mass within ETGs has been an ongoing process. 
Early analytic work involved the use of dynamical formalisms such as those of the Jeans equations \citep{Jeans22}, later implemented by \citet{Gunn77} and \citet{Binney82}, and analysing the effects of violent relaxation \citep{LyndenBell67, Shu78}. Assumptions of spherical symmetry and isotropy resulted in radial density distributions of the form $\rho(r) \propto r^{\gamma_{\rm tot}}$, with $\gamma_{\rm tot}\,=\,-2$, described as being `isothermal' for the combined distribution of stars and dark matter in a galaxy. 

Dark matter and stellar mass distributions have been found to interact with each other through processes such as dissipational collapse, in which the infall of baryonic matter compresses the dark matter halo \citep{Blumenthal86}. This study concluded that as a result of such interactions, it was unlikely that galaxies displayed purely isothermal profiles, as was suggested by earlier analytic studies. 

More recently, cosmological hydrodynamic simulations of galaxy formation have found that variations exist between the total mass density slopes ($\gamma_{\rm tot}$) values of galaxies and their individual formation histories. \cite{Remus13} utilised zoom-in cosmological simulations to find that more massive galaxies have a tendency toward shallower total-mass density slopes (over the stellar mass range $10^{11} < M_*/{\rm M}_{\odot} < 10^{12}$), with the flattening of the slope being driven by the number of dry merger events. 
Additionally, total-mass density slopes also correlated with the fraction of stars formed within the galaxy, with a shallower slope corresponding to a lower in-situ fraction. 
\citet{Remus17} found a tight correlation between $\gamma_{\rm tot}$ and central dark matter fractions, with the significance of the correlation depending strongly on the implementation of AGN (Active Galactic Nuclei) feedback. 
These results show that a galaxy's mass distribution is expected to be imprinted by the events of its assembly history. 

The total mass density profiles for EAGLE galaxies were analysed by \citet{Schaller15a}, who did not fit a total mass density slope to their profiles. It was instead noted that the galaxies are dark matter-dominated at all radii, and that the total mass density slopes are generally well-described by a \citet{Navarro96} profile, which does not have a constant $\gamma_{\rm tot}$ with radius. 

An increasing number of techniques have been developed that allow the total mass distribution to be inferred from observational data, enabling theoretical predictions of how mass assembles  in ETGs to be better tested. 
Early observational results analysing total-mass density profiles of galaxies utilised strong gravitational lensing, in which the mass distribution of the central region could be probed for the most massive galaxies in the Universe \citep{Auger10, Barnabe11, Sonnenfeld13}. These studies measured a small scatter in the total-mass density slopes of their galaxy samples, with on average a slightly `super-isothermal' total-mass density profile (i.e. steeper than $-2$). Furthermore, a subtle trend was recovered indicating that the most massive galaxies have shallower profiles, a trend also identified in a recent lensing study by \citet{Shu15}. Different techniques are required to probe total-mass density slopes for galaxies with lower masses as they do not have enough mass to produce strong lenses. These include dynamical modelling, either from central velocity dispersion values \citep[as done for a large sample of galaxies by][]{Tortora14}, or in greater detail with central 2D kinematics \citep[for example ][]{Poci17}. These studies all consistently measure slopes that have on average $\gamma_{\rm tot} < -2$. Recently, \citet{Lyskova18} combined both lensing and galaxy kinematics to provide additional constraints on the mass distributions of galaxies, and yielded total mass slopes that are also consistent with the aforementioned studies. Additionally, a semi-empirical approach by \citet{Shankar17} predicted slopes of $\gamma_{\rm tot} \sim -2.2$, becoming shallower at masses of $M_* > 10^{11.5}\rm M_{\odot}$. 

While the total mass density slope is a feature of galaxies that is of key interest, dynamical modelling of observed galaxies have focused on many aspects of galaxy structure and formation. These include an analysis of the stellar mass-to-light ratio and the initial mass function of galaxies \citep[e.g.][]{Thomas11, Posacki15, Tortora16, Li17, Oldham18}, the orbital anisotropy within galaxies \citep[e.g.][]{Magorrian01}, intrinsic shapes of galaxies \citep[e.g.][]{vandenBosch08}, decompositions of baryonic and dark matter mass within galaxies \citep[e.g.][]{Thomas11, Tortora12, Zhu16}, and the nature of dark matter itself \citep[e.g.][]{Salinas12, Chae15, Samurovic16}. While we do not directly address such goals within this work, their acknowledgement is key in building a complete picture of the successes of galaxy dynamical modelling.

The local dark-matter fraction of a galaxy varies with radius. Within the central region, stellar mass dominates. At a few effective radii the contributions of stellar mass and dark matter mass become comparable, and at larger radii the dark matter dominates \citep[for example][]{Remus13, Laporte15, Dutton16}. The aforementioned observational studies therefore all probe mass distributions in regions dominated by stellar mass \citep[as shown through dynamical modelling of the ATLAS$^{\rm 3D}$ survey by][]{Cappellari13}. To gain an understanding of the structure of the dark matter halo, it is important to analyse the dynamics over greater radial ranges in galaxies. 
This was done by \citet{Cappellari15}, combining central-IFU (Integral Field Unit) ATLAS$^{\rm 3D}$ data and multi-slit data for 13 of the most massive galaxies in the SLUGGS survey \citep{Brodie14}, whose data extend to radii of $\sim4\,-\,6R_e$. For this sample of $M_*>10^{11}M_{\odot}$ galaxies, the average total-mass density slope was measured to be $\gamma_{\rm tot} = -2.19 \pm 0.03$, consistent with the results for the central regions of galaxies, indicating that the total-mass density slopes of massive ETGs remain roughly constant out to radii of $\sim 4\,R_e$.

At low stellar masses ($M_*<10^{11}M_{\odot}$), ETGs look quite different from their older, more massive siblings. While massive ETGs are predominantly `slow rotator' ellipticals, the fraction of lenticular (S0) and centrally `fast rotating' elliptical galaxies is much larger at lower stellar masses \citep[for example][amongst many others]{Kormendy82, Davies83, Nieto89, Kormendy09, Emsellem11}. While massive elliptical galaxies are generally described as being the result of multiple mergers in the `two-phase formation' scenario \citep{Oser10, Johansson12}, there are several formation mechanisms which have been proposed to describe the formation of lenticulars. Simulations have shown that mergers are able to produce S0-like galaxies \citep[e.g.][]{Bekki98, Bois11, Tapia14, Querejeta15}, while the presence of `pseudobulges' (versus the classical bulges seen in elliptical galaxies) has been interpreted as an indication that S0 galaxies have undergone a more secular history \citep{Kormendy04, Laurikainen06}. Such `secular' processes result in the quenching of disc-like galaxies through  mechanisms such as galactic winds \citep{Faber76}. Additionally, environmental processes have been proposed to produce quenched S0s, such as ram pressure stripping \citep{Gunn72}, strangulation \citep{Larson80}, thermal evaporation \citep{Cowie77} and galaxy harrassment \citep{Moore96, Moore99}.

The SLUGGS\footnote{SAGES Legacy Unifying Globulars and GalaxieS, \url{http://sluggs.swin.edu.au}} Survey \citep{Brodie14} has utilised the Keck telescope to collect spectral data of stellar and globular cluster light from 25 early-type galaxies out to radial distances of up to 10 $R_e$. In addition to the large radial extent of the data, another advantage of the SLUGGS survey (over other surveys with larger samples, such as ATLAS$^{\rm3D}$ \citealt{Emsellem11}, SAMI \citealt{Bryant15} or CALIFA \citealt{Sanchez12}) is the higher $\sigma$ resolution of $\sim24$ km s$^{-1}$. As a result, this survey has been able to measure kinematics and metallicities out to larger radii than previous spectroscopic studies. 
This also allows for the extended structure of dark-matter haloes to be derived with mass-modelling techniques \citep[e.g.][]{Alabi17}. 

In this paper, we use updated SLUGGS and post-SLUGGS data combined with ATLAS$^{\rm 3D}$ \citep{Emsellem11} data to extend the \citet{Cappellari15} sample to lower stellar masses, in order to better understand whether the total mass distributions vary at larger radii, and to help discriminate between the many potential formation and evolution mechanisms that mould early type galaxies at lower stellar masses. 

In order for total-mass density slopes to be used to identify different formation histories, it is useful to extract predictions from cosmological simulations, which are able to connect present-day total-mass density slopes of galaxies to their histories. For this, we present predictions from the \textit{Magneticum Pathfinder} simulations (Dolag et al. in prep.). As an additional comparison, we also include total-mass density slope measurements made over a broad stellar mass range for galaxies in the EAGLE simulations \citep{Schaye15}. 

The structure of this paper is as follows.
Section \ref{sec:Data} describes the observed and simulated data used in this paper, Section \ref{sec:DynamicalModelling} details the modelling techniques we utilise, where we describe the techniques used to calculate $\gamma_{\rm tot}$ values for individual galaxies. The results from our dynamical modelling are presented in Section \ref{sec:ModellingResults}, and we discuss the modelling results themselves in Section \ref{sec:ModellingDiscussion}. The manner in which we extract $\gamma_{\rm tot}$ values from simulated data is outlined in Section \ref{sec:SimulationsGamma}.  The scientific results are discussed in Section \ref{sec:ScientificContext}, future work is described in Section \ref{sec:FutureWork}, and we present our conclusions in Section \ref{sec:Conclusion}. 


\section{Data}
\label{sec:Data}

\subsection{Observations}

We utilise data from the SLUGGS Survey \citep{Brodie14} taken with the DEIMOS instrument on the Keck telescope. The galaxies focused on in this work are NGC 1052, NGC 2549, NGC 2699, NGC 4459, NGC 4474, NGC 4551, NGC 5866 and NGC 7457\footnote{Although NGC 7457 was analysed by \citet{Cappellari15}, we include this galaxy in our central sample because additional observations of this galaxy have since been made. These new data were presented by \citet{Bellstedt17a}. }. Three of the galaxies analysed in this study, NGC 1052, NGC 2549 and NGC 2699, do not fall within the original SLUGGS sample \citep[as summarised by][]{Brodie14}. For these galaxies, the data were collected in broadly the same manner as for the main SLUGGS survey, using central slits to collect stellar light out to 2-3 effective radii ($R_e$), and additional slits in the galaxy outskirts to collect spectra for globular clusters \citep[a catalogue of SLUGGS globular clusters was published by][]{Forbes17b}. To maximise the central slits focussing on stellar light, we utilise the SuperSKiMS method outlined by \citet{Pastorello16}. We do not use the globular cluster spectra in this paper, as an adaptation of the JAM method to discrete tracers is beyond the scope of this work. We also present an updated dataset for NGC 5866, for which two extra DEIMOS masks are included, in addition to the previously published dataset (consisting of one DEIMOS mask) by \citet{Foster16}. All galaxies except NGC 1052 are part of the ATLAS$^{\rm 3D}$ sample \citep{Cappellari11a}. 

\begin{figure*}
	\centering
	\includegraphics[width=180mm]{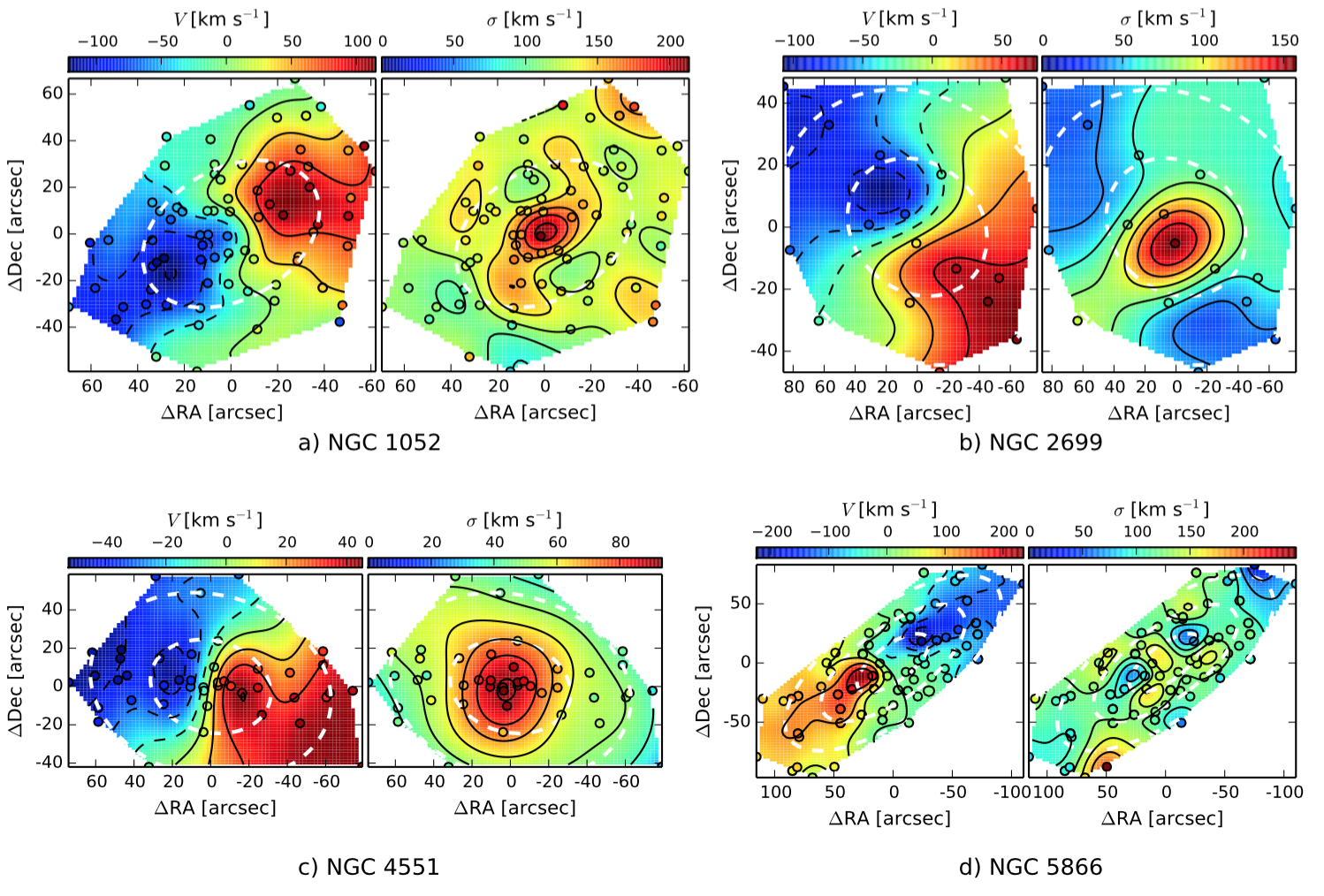}
	\caption{2D kinematic maps for NGC 1052, NGC 2699, NGC 4551 and NGC 5866. The slit positions are indicated as small circles, and the iso-velocity contours are presented as black lines. The interpolation of each kinematic field is produced through the Kriging technique. An indication of the spatial extent of each map is given by the 1 and 2 $R_e$ dashed white ellipses. The colour bar for each map shows the mean velocities/velocity dispersions, from lowest values (blue) to highest values (red). Previous maps for NGC 5866 have been published by \citet{Foster16}. This dataset includes observations from two additional DEIMOS masks. For each of the kinematic maps, the orientation is north up, and east to the left.  }
	\label{fig:Kinematics}
\end{figure*}

Table \ref{tab:ObservationalSummary} summarises the observations contributing to our datasets. Kinematic maps for galaxies NGC 2549, NGC 4459, NGC 4474, and NGC 7457 were presented by \citet{Bellstedt17a}, whereas for NGC 1052, NGC 2699, NGC 4551 and NGC 5866 the kinematic maps are presented here in Fig. \ref{fig:Kinematics}. These maps are generated through application of the Kriging interpolation technique, as outlined by \citet{Pastorello14}.

\begin{table}
	\centering
	\caption[Observation Parameters]{Observations for NGC 1052, NGC 2699, NGC 4551 and NGC 5866. See table 2 of \citet{Bellstedt17a} for the observational summary for the remaining galaxies.   }
	\label{tab:ObservationalSummary}
	\begin{tabular}{@{}r ccc}
		\hline
		\hline
		Mask Name & Obs. Date & Exp. Time & Seeing     \\
		&  & (sec) &  (arcsec)   \\
		(1)  & (2) & (3) & (4) \\
		\hline
		\multicolumn{4}{l}{\textit{Existing data:}} \\
		1N5866B & 2013-04-11 & 3600 & 0.7   \\
		\multicolumn{4}{l}{\textit{New data:}} \\
		1N1052 & 2016-11-25 & 7200 & 1.0   \\	
		2N1052 & 2016-11-25 & 7200 & 1.3   \\
		3N1052 & 2017-01-28 & 6762 & 0.7   \\
		4N1052 & 2017-01-29 & 6720 & 0.6   \\
		1N4551 & 2016-03-11 & 7200 & 0.8 \\
		2N5866 & 2017-04-28 & 5400 & 0.6 \\
		3N5866 & 2017-04-29 & 11400 & 1.0 \\
		1N2699 & 2016-11-25 & 12600 & 1.5 \\
		2N2699 & 2016-11-26 & 7200 & 1.5 \\
		2N4551 & 2017-01-28 & 7200 & 0.7 \\
		3N4551 & 2017-01-29 & 7200 & 0.6 \\
		\hline
	\end{tabular}
\end{table}

In the past, NGC 1052 has been suspected to be an example of a prolate galaxy, in which the galaxy rotates along the minor axis, instead of the major \citep{Schechter79}. Our 2D kinematic maps in Fig. \ref{fig:Kinematics} show that NGC 1052 is in fact an oblate galaxy, rotating along its major axis. The left panel of part a) Fig. \ref{fig:Kinematics} displays clear rotation along the major axis of the galaxy, with a rotational velocity of $\sim120$ km s$^{-1}$, and central velocity dispersion of $\sim200$ km s$^{-1}$. 

Part b) of Fig. \ref{fig:Kinematics} presents the 2D kinematic maps for NGC 2699. It also displays clear rotation along the major axis, with a central peak in velocity dispersion of $\sim150$ km s$^{-1}$. We note that the sampling of observational points (as indicated by the small circles on the kinematic maps) is sparser for this galaxy than for the other three as a result of the small angular size of the galaxy. 

Our data, and the kinematic maps, for NGC 4551 are shown in part c) of Fig. \ref{fig:Kinematics}. As for NGC 1052 and NGC 2699, the velocity map shows a regular rotation along the major axis of the galaxy, albeit with a significantly lower rotational velocity of $\sim45$ km s$^{-1}$. NGC 4551 also has a large central $\sigma$ peak of $\sim90$ km s$^{-1}$.

Kinematic maps for NGC 5866 are shown in part d) Fig. \ref{fig:Kinematics}. While this galaxy displays rotation along the major axis just like the other galaxies of this study, this rotation declines at larger radii. This hints at the presence of an embedded disc within the galaxy. Our data for this galaxy extend to $\sim3\,R_e$, as indicated by the dashed white ellipses in both panels of Fig. \ref{fig:Kinematics} part d).

A set of basic properties for these galaxies is outlined in Table \ref{tab:GalaxyProperties}. 
\begin{table*}
	\centering
	\caption[Galaxy Properties]{Galaxy Properties}
	\label{tab:GalaxyProperties}
	\begin{threeparttable}
		\begin{tabular}{@{}c ccccccccccc}
			\hline
			\hline
			Galaxy & R.A. & Dec. & $M_K$ & log($M_*$) & Dist  & $\sigma_{\rm kpc}$ & $R_e$ & Morph. & P.A. & $\epsilon$ & $V_{sys}$ \\
			(NGC) & (h m s) & (d m s) & (mag) & (log($M_{\odot}$)) & (Mpc)  & (km s$^{-1}$) & (arcsec) &  & (deg) &  & (km s$^{-1}$)  \\
			(1) & (2) & (3) & (4) & (5)  & (6) & (7) & (8) & (9) & (10) &(11) & (12)\\
			\hline
			1052 &  02 41 04.8 & $-$08 15 21 & $-$22.58 & 11.02 & 18.8 & 197 & 21.9 & E3-4/S0 & 120.0$^{\dagger}$ & 0.31$^{\dagger}$ & 1510$^{\dagger}$ \\
			2549 &  08 18 58.3 & +57 48 11 & $-$22.43 & 10.28 & 12.3 & 141$^{\triangleleft}$ & 14.7 & S0 & 179.5$^{\star}$ & 0.69$^{\star}$ & 1051$^{\ddagger}$ \\
			2699 &  08 55 48.8 & $-$03 07 39 & $-$22.72$^{\ddagger}$ & 10.40 & 26.2 & 131$^{\triangleleft}$ & 11.5$^{\ddagger}$ & E & 46.8$^{\star}$ & 0.14$^{\star}$ & 1868$^{\ddagger}$ \\
			4459 &  12 29 00.0 & +13 58 42 & $-$23.89 & 10.98 & 16.0 & 170 & 48.3 & S0 & 105.3 & 0.21 & 1192 \\
			4474 &  12 29 53.5 & +14 04 07 & $-$22.27 & 10.23 & 15.5 & 88 & 17.0 & S0 & 79.4 & 0.42 & 1611 \\
			4551 &  12 35 37.9 & +12 15 50 & $-$22.18 & 10.24 & 16.1 & 97$^{\triangleleft}$ & 13.8 & E & 75.0$^{\star}$ & 0.25$^{\star}$ & 1191$^{\ddagger}$ \\
			5866 &  15 06 29.5 & +55 45 48 & $-$24.00 & 10.83 & 14.9 & 163 & 23.4 & S0 & 125.0 & 0.58 & 755 \\
			7457 &  23 00 59.9 & +30 08 42 & $-$22.38 & 10.13 &  12.9 & 74 & 34.1 & S0 & 124.8 &  0.47& 844 \\
			\hline
		\end{tabular}
		\begin{tablenotes}
			\item Notes: 
			(1) Galaxy NGC number. 
			(2) Right ascension and
			(3) Declination (taken from the NASA/IPAC Extragalactic Database).
			(4) $K$-band magnitude.
			(5) Total stellar mass calculated from $3.6\,\mu{\rm m}$ magnitude.
			(6) Distance.
			(7) Central velocity dispersion within 1 kpc. 
			(8) Effective radius.
			(9) Morphology.
			(10) Photometric position angle. 
			(11) Ellipticity. 
			(12) Systemic velocity.
			Columns (5) and (8) are taken from \citet{Forbes17}, except for NGC 2699 which was taken from \citet{Cappellari11a} (stellar mass calculated from the $K$-band magnitude, assuming $(M/L)_*=1$).
			Values from columns (7), (9)--(12) are taken from \citet{Brodie14}, except for the following entries: [$^{\ddagger}$] \citet{Cappellari11a}; [$^{\triangleleft}$] \citet{Cappellari13}; [$^{\star}$] \citet{Krajnovic11}; [$^{\dagger}$] NED.
			
		\end{tablenotes}
	\end{threeparttable}
\end{table*}

\subsection{Simulations}

\subsubsection{Magneticum}

We compare our results with those of simulated ETGs from the \textit{Magneticum Pathfinder}\footnote{\url{www.magneticum.org}} (Dolag et al. in prep.) cosmological hydrodynamic simulation.
These simulations were performed with an updated version of the TREE-SPH code {\sc gadget3}, as described in detail in \citet{Hirschmann14} and \citet{Teklu15}, most importantly including feedback from AGN \citep{Springel05, Fabjan10, Hirschmann14}, as well as stellar feedback \citep{Springel05} and metal enrichment from SN Ia, SN II and AGB mass loss \citep{Tornatore04,Tornatore07}. We selected ETGs from a box with a volume of (48$h^{-1}$\,Mpc)$^3$ and a particle resolution of $m_\mathrm{DM}=3.6\times10^7h^{-1}\,\rm M_\odot$ and $m_\mathrm{gas} =
7.3\times10^6h^{-1}\,\rm M_\odot$, with every gas particle spawning up to 4 stellar particles, resulting in a stellar particle resolution of $m_\mathrm{star}\approx2\times10^6h^{-1}\,\rm M_\odot$. The gravitational softening at $z=0$ for the stellar particles is $\epsilon_*=0.7h^{-1}$\,kpc.  A WMAP7 \citep{Komatsu11} $\Lambda$CDM cosmology was adopted with $h=0.704$, $\Omega_\Lambda=0.728$, $\Omega_\mathrm{bar} = 0.0451$ and $\sigma_8 = 0.809$.
The initial power spectrum has a slope of $n_s = 0.963$.

The ETGs were classified using the same criterion as \citet{Remus17}, i.e. based on their specific angular momentum. 
In particular, for each star inside $0.1R_\mathrm{vir}$ of a galaxy, the circularity $\epsilon_\mathrm{circ}\equiv j_z/j_\mathrm{circ}$ was calculated, with $j_z$ the specific angular momentum of each particle with respect to the principal axis of inertia of the galaxy and $j_\mathrm{circ}$ the specific angular momentum of the particle if it were on a circular orbit at the same radius. 
If more than $40\%$ of the star particles within a galaxy have circularities within $-0.3 \leqslant \epsilon_\mathrm{circ} \leqslant 0.3$, and the galaxy itself has a cold gas fraction within the stellar half-mass radius of $m_{\rm gas}/m_*<3\%$, we consider it to be an ETG. For more details on the choice of these values see \citet{Teklu15}. 
Differently to the work presented in \citet{Remus17}, here we consider all galaxies (187) with masses above $M_* > 10^{10}M_\odot$ at $z=0$ for which in-situ fractions could be obtained, in order to include lower mass ETGs for comparison with our observational data. 
The lower mass limit is a good match  to that of the SLUGGS survey galaxies, and is chosen such that the galaxies have enough particles for the equal-particle bins. 
This ensures that the spatial resolution is sufficient to allow stellar equal-particle bins with a size of 80 particles or more per bin per galaxy to be used. 
All fits and calculations of the parameters taken from the simulation were determined with the same method as presented in \citet{Remus17}.

\subsubsection{EAGLE}

As an additional point of comparison to simulations, we also measure the density profiles and their slopes from ETGs in the EAGLE cosmological hydrodynamic simulations \citep{Schaye15}.  The simulations were run with a private variation of {\sc gadget3}, which included a modified SPH scheme \citep[see][]{Schaller15b}.  $\Lambda$CDM parameters based on \citet{PlanckXVI} were assumed ($h=0.6777$,  $\Omega_{\Lambda}=0.693$, $\Omega_{\rm bar}=0.048$, $\sigma_8=0.8288$  $n_s = 0.9611$).  Formation of star particles (whereby single gas particles are converted into single star particles) was based on local gas pressure \citep{Schaye04,Schaye08}, where the stellar evolution of each particle was followed as a single stellar population \citep{Chabrier03,Wiersma09}, resulting in stochastic thermal stellar feedback \citep{Dalla12}.  Seeded black holes \citep{Springel05} accreted nearby gas particles, driving stochastic thermal AGN feedback \citep[detailed in][]{Schaye15}.  The simulations were calibrated at $z=0.1$ to match the observed galaxy stellar mass function and stellar size--mass relation \citep[see][]{Crain15}.  Haloes and subhaloes were identified with {\sc subfind} \citep{Springel01,Dolag09}.  Both the halo and particle data for these simulations are publicly available\footnote{\url{http://icc.dur.ac.uk/Eagle/database.php}} \citep{McAlpine16}.

We include results from the main EAGLE run containing $1504^3$ initial particles of dark matter and gas each, with respective masses $m_{\rm DM} \! =  6.57  \times  10^6 h^{-1}\,\rm M_{\odot}$ and $m_{\rm gas} =  1.22  \times  10^6 h^{-1}\,\rm M_{\odot}$, in a periodic box of volume (67.77$h^{-1}$\,Mpc)$^3$, with a softening length of $474.4h^{-1}$\,pc at $z<2.8$. We additionally address results from the high-resolution, recalibrated simulation (eight-times smaller particles masses, half the softening length, one quarter the box length).  Because the sub-grid models in EAGLE are resolution-dependent, to meet the same observational constraints for the high-resolution run, the density dependence of stellar feedback was altered and the temperature boost to gas from AGN feedback was increased \citep[see][]{Schaye15}.
		
We extract ETGs from EAGLE with total stellar mass above $10^{10}\,{\rm M}_{\odot}$ at $z=0$, using the same gas fraction and circularity cuts as used for \emph{Magneticum}.  We exclude all satellite galaxies, and only include centrals that are dynamically relaxed: specifically those where the centres of potential and mass are separated by less than $0.07 R_{\rm 200c}$\footnote{$R_{\rm 200c}$: the radius within which the average density is 200 times the \textit{critical} density of the Universe.} \citep[see][]{Maccio07,Schaller15a}. This results in a final sample of 288 galaxies. 

\section{Dynamical Modelling}
\label{sec:DynamicalModelling}

Since we can only directly measure properties of the luminous component of each galaxy, we require dynamical modelling to gain an understanding of the total mass distribution. 
Jeans Anisotropic MGE (JAM) modelling \citep{Cappellari08} is a computationally efficient, openly available mass-modelling technique that utilises the Jeans equations \citep{Jeans22} and a simple, 2D parametrisation of galaxy light to obtain dynamical models of individual galaxies.
The JAM technique and our implementation thereof are outlined in the following section. 

For readers not interested in a detailed description of the JAM modelling method and discussion, we suggest skipping directly to Section \ref{sec:ScientificContext}. 


\subsection{Jeans Anisotropic MGE Modelling}

JAM modelling solves the axisymmetric Jeans equations \citep{Jeans22} for the kinematics of a galaxy based on a parametrisation of the galaxy mass distribution. Given a model, Markov Chain Monte Carlo (MCMC) can be used to determine the parameters which give the best fit to the observed kinematics. 
This method allows the dynamical parameters, such as dynamical mass, total-mass density profiles, anisotropy, and inclination, to be estimated for galaxies that are axisymmetric. 

The observed kinematics are fed into the JAM model in the form of the root-mean-square velocity:  $V_{\rm rms} = \sqrt{V^2 + \sigma^2}$. This form is selected as it uses information from both the velocity and velocity dispersion fields of the galaxy in a single kinematic map. 
The stellar light distribution is implemented as a set of fitted Gaussians as an input into JAM. 
These Gaussian components are derived utilising a Multi-Gaussian Expansion \citep[MGE,][]{Cappellari02}, which fits the 2D distribution of stellar light. 

When applying JAM modelling to galaxies of the Illustris simulations, \citet{Li16} determined that JAM produced total mass profiles of the simulated galaxies with a mean error of 26 percent, indicating that although the outputs from JAM are not perfect, the parameters do recover the underlying mass distribution. 

\subsubsection{Input Mass Models}

In this study, we implement two models separately with the aim of determining which better describes the galaxies. 
Because of its simplicity, we implement a double power law model (as has been done by many dynamical modelling and strong gravitational studies), and an additional model in which the stellar and dark matter mass are parametrised separately. The first model tests how well the power law mass distribution can recover the galaxy kinematics, while the second challenges the power law assumption by testing whether an increase in the mass distribution flexibility will improve the quality of the model kinematics. 
We outline each of these in the following.\linebreak

\paragraph{Model 1}
The distribution of \textit{total} mass here is assumed to be in the form of a generalised \citet{Navarro96} (NFW) profile, given by:
\begin{equation}
\rho(r) = \rho(r_s) \left(\frac{r}{r_s}\right)^{\gamma}\left(\frac{1}{2}+\frac{r}{2r_s}\right)^{-\gamma-3},
\label{eqn:gNFW}
\end{equation}
\citep[as presented in][]{Hernquist90, Navarro96, Merritt06} in which the power law slope within the scale radius $r_s$ is given by the free parameter $\gamma$, and the outer slope is fixed at $-3$. 

The density at the scale radius $\rho(r_s)$ has the function of scaling the amount of mass present in the profile. More mass can be added to a profile by not only increasing $\rho(r_s)$ however, but also by decreasing $\gamma$ to steepen the profile. 
As a result, $\rho(r_s)$ and $\gamma$ are not unique parameters, and can be seen to covary. Hence, MCMC displays difficulty in adequately constraining these parameters. In order to improve the constraint, we rearrange equation \ref{eqn:gNFW} such that the scaling density is defined at a radial point $x$ at which we have observational data, rather than at the scale radius, as done by \citet{Mitzkus17}. This rearranged form of the density profile is given as:
\begin{equation}
\rho(r) = \rho(x)\left(\frac{r}{x}\right)^{\gamma}\left(\frac{r_s + r}{r_s + x}\right)^{-\gamma - 3},
\label{eqn:PowerLaw}
\end{equation} 
where we select $x$ to be 1 kpc. The value of $x$ is arbitrary and can be defined at any radius, although we note that a larger separation of $x$ from $r_s$ provides a better constraint. Throughout this work, we fix the scale radius to 20 kpc. We discuss in Appendix \ref{sec:FreeScaleRadius} why letting the scale radius be a free parameter does not improve our method, and that fixing it is valid. 

The free parameters in this model are the inclination $i$, the anisotropy $\beta$, the total-mass density at 1 kpc $\rho(1)$, and the inner total-mass density slope $\gamma$. The parameter $\beta$ describes the anisotropy of the velocity distribution, given by $\beta = 1-\overline{v_{\theta}^2}/\overline{v_{r}^2}$ \citep[eq. 4-53b of][]{Binney87}. A positive $\beta$ is an indicator of \textit{radial} anisotropy, while a negative value is an indicator of \textit{tangential} anisotropy. Isotropic systems have $\beta = 0$.
The prior parameter ranges are as follows: 
$i_{\text{min}} < i \leq 90^{\circ}$, 
$-0.5 < \beta < 0.5$, 
$-2.5 < \gamma < -1.5$ , and 
$-1 < \log(\rho(1)) < 3$ (where $\rho(1)$ is measured in $\rm M_{\odot}\rm pc^{-3}$). 
The minimum inclination $i_{\text{min}}$ is calculated assuming that the intrinsic axial ratio of the galaxy is 0, such that $i_{\text{min}} = \cos^{-1}(q)$, where $q$ is the axial ratio of the most flattened Gaussian component fitted to the stellar light of the galaxy \citep{Cappellari08}. 

The inner total mass density slope $\gamma$ presented in Eqn. \ref{eqn:gNFW} is not identical to the fitted total mass density slope used to describe the global potential of a galaxy, denoted by $\gamma_{\rm tot}$. In our implementation, $\gamma_{\rm tot}$ is not directly a free parameter, and is instead fitted to the retrieved total mass density profile of each galaxy. 
The total mass density slope $\gamma_{\rm tot}$ is calculated by fitting a power-law slope to the $0.1-4\,R_e$\footnote{For some of the smaller galaxies in our sample, SLUGGS data do not extend to the full $4\,R_e$. For these galaxies, we fit the profiles to the maximum radial extent of our data. This is discussed in more detail in Section \ref{sec:FinalValueSelection}. } radial range of the profile. This value is slightly steeper than the free parameter $\gamma$.

\paragraph{Model 2}
In this model, we describe the total mass distribution as the sum of a stellar mass distribution and a spherical DM halo in the form of a generalised NFW distribution, as given by equation \ref{eqn:PowerLaw} (except that rather than using the symbol $\gamma$, we use the symbol $\alpha$ to represent the inner slope of the dark matter component). We select this parametrisation over a simple NFW profile as it conveys the effects of baryonic interactions \citep[for example][]{Navarro96b, Governato12, Ogiya14, Peirani17, Peirani18}. The distribution of stellar mass is taken from the input parametrisation of the observed luminosity. 

In this parametrisation of the galaxy mass, not only do we recover the parameters of the generalised NFW distribution describing the dark matter, we also recover the stellar mass component by scaling the stellar light by a stellar mass-to-light ratio. This accounts for stellar population variations between galaxies. 
Note, we assume that this $(M/L)_*$ is constant with radius (see \citealt{Poci17, Mitzkus17} for examples in which radial variations in $(M/L)_*$ are considered).  
The effect of having a constant $(M/L)_*$ with radius is that any variation in stellar mass that would be due to a varying $(M/L)_*$ is instead attributed to the dark matter halo. This means that the exact decomposition between baryonic and dark matter mass may suffer from the bias caused by this assumption. The derived \textit{total} mass density slope, however, should not be notably affected by this assumption.
The free parameters in this model are the inclination $i$, the anisotropy $\beta$, the dark matter density at 1 kpc $\rho(1)$, the dark matter inner density slope $\alpha$, and the mass-to-light ratio $(M/L)_*$. As with model 1, we fix the scale radius of the dark matter profile to 20 kpc. 

In order to make a measurement of the \textit{total} mass density slope, we fit a slope to the combined stellar and dark matter mass. 
The 1D radial profiles (whether stellar, dark matter or total) input into JAM are line-of-sight projected quantities. Therefore, before we can fit a slope to the profiles, they must first be deprojected into 1D intrinsic profiles. This is achieved by applying the equation from footnote 11 of \citet{Cappellari15} to the projected MGE:
\begin{equation}
\rho(r) = \sum_{j=1}^M \frac{ M_j \exp{[-r^2 / (2\sigma_j^2)]} \text{erf}[r\sqrt{1-q_j^{\prime\,2}} / (q^{\prime}_j\sigma_j\sqrt{2})] }{4\pi\sigma_j^2r\sqrt{1-q_j^{\prime\,2}}}.
\end{equation}
Here, $j$ indicates the individual Gaussians that make up the MGE describing the projected 2D distribution, $M_j$ is the total mass of each MGE (given by $M_j = 2\pi q^{\prime}_j\Sigma_j \sigma_j^2$), and $\sigma_j$ the dispersion. We also note that $q_j^{\prime}$ is the \textit{intrinsic} axial ratio of each Gaussian (still assuming axisymmetry), as opposed to the projected axial ratio $q_j$. This is given by $q^{\prime}_j = \sqrt{q_j^2 - \cos^2{i}} / \sin{i}$. 

The value for $\gamma_{\rm tot}$ in this implementation is established by fitting a power law to the 1D stellar + dark matter mass profiles over a specified radial range. We discuss in Section \ref{sec:PowerLawAssumption} the effect of selecting different radial ranges for this fit. \linebreak
\begin{table*}
	\centering
	\caption[JAM Input]{Black hole masses and MGEs used for JAM modelling. }
	\label{tab:JAMInputParameters}
	\begin{tabular}{@{}c | ccc}
		\hline
		\hline
		Galaxy & $M_{BH}$ & $M_{BH}$ Source & MGE Source \\
		& ($\times10^{7} M_{\odot}$) & &  \\
		(1) & (2) & (3) & (4)  \\
		\hline
		NGC 1052 & 9.4 & \citet{Beifiori09} & This work\\
		NGC 2549 & 1.4 & \citet{McConnell13} & \citet{Scott13} \\
		NGC 2699 & 1.2 & $M_{\rm BH} - \sigma$ Scaling Relation  & \citet{Scott13} \\
		NGC 4459 &7.0 & \citet{Woo13} & \citet{Scott13}\\
		NGC 4474 & 0.4 & \citet{Gallo10}  & \citet{Scott13}\\
		NGC 4551 & 0.63 & \citet{Gallo10} & \citet{Scott13}\\	
		NGC 5866 & 8.26 & \citet{Saikia15} & \citet{Scott13}\\
		NGC 7457 & 4.1 & \citet{Hartmann14} & \citet{Cappellari06} \\
		\hline
	\end{tabular}
\end{table*}

For both models, we include a black hole mass ($M_{\rm BH}$) within the mass model for JAM modelling. For galaxies that do not have published black hole masses, we utilise the $M_{\rm BH} - \sigma$ scaling relation as published by \citet{Graham13} ($\log(M_{\rm BH}) = 8.20 + 6.08\log(\sigma/200\,{\rm km\,s^{-1}})$) to estimate $M_{\rm BH}$. As discussed in Section \ref{sec:PreparationOfData}, the accuracy of the black hole masses do not affect our results, and therefore we can use such approximations. The black hole masses used are indicated in Table \ref{tab:JAMInputParameters}, as are the literature sources of MGEs used within JAM for each galaxy to describe the distribution of stellar light. 

\subsubsection{MCMC Implementation}
\label{sec:MCMCImplementation}

To determine the value of the parameters that best describe the galaxy kinematics, we wrap JAM within an MCMC process, to explore the full parameter space for each model. 
We use \texttt{emcee} \citep{Foreman-Mackey13} to run MCMC for our work. The MCMC code used within this paper has been set up with 200 walkers each doing 4000 steps, after rejecting an additional 1000 burn steps. 

For MCMC to determine the best-fitting parameters, we model the likelihood function as a Gaussian, where we use our dynamical model to predict $V_{\rm rms}^{\prime}$, which is compared to the observed $V_{\rm rms}$.
Usually this is done as a single calculation for a single dataset. 
Both the SLUGGS and ATLAS$^{\rm 3D}$ datasets were used to run the JAM models for each galaxy for which ATLAS$^{\rm 3D}$ data are available. 
Since the distribution of points and typical uncertainties are different for the two datasets, we have instead chosen to run JAM on each dataset separately. We thereby determine a separate likelihood for each dataset, and implement hyperparameters as additional free parameters to effectively ``weight" the datasets, as outlined by \citet{Hobson02}. 
We then combine the log likelihood of each dataset as given by:
\begin{equation}
\ln \Lagr_{\rm total} = \ln \Lagr_{\rm SLUGGS} + \ln \Lagr_{\rm ATLAS3D},
\label{eqn:totalLikelihood}
\end{equation}
where the individual likelihoods are calculated as:
\begin{equation}
\ln \Lagr = -\frac{1}{2}\sum_{i=0}^{n}\left[\ln\left(\frac{(2\pi)\Delta V_{{\rm rms}, i}^2}{\omega}\right) + \omega\left( \frac{V_{{\rm rms}, i}-V_{{\rm rms}, i}^{\prime}}{\Delta V_{{\rm rms}, i}}\right)^2 \right],
\label{eqn:weights}
\end{equation}
for the $n$ datapoints in each dataset. Here, $V_{\rm rms}$ represents the input kinematic values, $V_{\rm rms}^{\prime}$ represents the corresponding model kinematic values, and $\Delta V_{\rm rms}$ represents the uncertainties in the input kinematics. The hyperparameter for the dataset is indicated by $\omega$, which we include as a free parameter in each galaxy for which we have both SLUGGS and ATLAS$^{\rm 3D}$ datasets.
A higher hyperparameter value $\omega$ increases the weight of the associated dataset, while a lower $\omega$ weakens the influence of the dataset by increasing the associated uncertainty. We note that the use of the hyperparameter method is applicable in this case because the SLUGGS and ATLAS$^{\rm 3D}$ datasets are independent. 
This technique is described in more detail by \citet{Hobson02}. 

The SLUGGS and ATLAS$^{\rm 3D}$ datasets had previously been combined by \citet{Cappellari15}, who did so by manually increasing the ATLAS$^{\rm 3D}$ uncertainties such that the central dataset did not dominate the overall likelihood calculation. We highlight that our hyperparameter implementation is an improvement over this method, as it is more statistically rigorous, and does not alter the original uncertainties.

The final value for each parameter is selected as the value from the MCMC iteration that maximises the likelihood, where asymmetric uncertainties are given as a 68.3\% ($1\,\sigma$) credible interval around the maximum likelihood, such that the upper and lower bounds have equal posterior probability density values. 
We use the \texttt{ChainConsumer} python package \citep{Hinton16} to generate these values from the \texttt{emcee} outputs.


\subsection{Preparation of data}
\label{sec:PreparationOfData}

As a result of the large spatial extent of the kinematic data, variation in sampling density, variations in signal-to-noise ratio and uncertainties, the input kinematic data need to be cleaned to ensure the best dynamic modelling by JAM. 
After the initial spectrum quality-check by eye, additional cleaning measures are implemented. 
To account for the relatively sparse sampling of data points in the SLUGGS data, we initially bi-symmetrise the data by duplicating the points across both the $x$- and $y$-axes \citep[as done in][]{Cappellari15}. 
We then apply a signal-to-noise threshhold of 20 to the data, and eliminate any remaining points with a $>100\%$ uncertainty in $V_{\rm rms}$.\footnote{$\Delta V_{\rm rms}$ is determined by adding the uncertainties of $v$ and $\sigma$ in quadrature.} We also account for the observed velocity dispersion offset between the two datasets by increasing the velocity dispersion of SLUGGS data by the offset amount (see next section for details). 
For the purpose of JAM modelling, we are interested in the global kinematic trends, and not any kinematic substructures that may be present. To best highlight these global kinematic trends, we apply the LOESS\footnote{LOcally weighted regrESSion} smoothing algorithm of \citet{Cleveland88} to the data as done in \citet{Cappellari15}, which extracts the underlying kinematic trends and eliminates smaller scale features resulting from either minor substructure or noise. 

We analyse the effect that each of these procedures has on the final $\gamma_{\rm tot}$ measurements, by running JAM with and without them, for both input models. Not accounting for the offset in velocity dispersion values between the two datasets (see Section \ref{sec:VelocityDispersionOffsetImpact}) results in a mean scatter of 0.057 for model 1 (0.053 for model 2), not smoothing the $V_{\rm rms}$ data results in a mean scatter of 0.040 (0.077), and when not smoothing or symmetrising the input data, the resulting mean scatter is 0.056 (0.074). We do not see any systematic effects, and none of these values are significant enough to affect our results. We also assess the scatter in $\gamma_{\rm tot}$ caused by not including a black hole mass, which is 0.012 (0.029). The fact that this scatter is so low (compared to the uncertainties) indicates that our results are not sensitive to inaccurate black hole mass measurements.

\subsubsection{Impact of velocity dispersion offset}
\label{sec:VelocityDispersionOffsetImpact}

It has been determined by a number of studies \citep{Foster16, Pastorello16, Bellstedt17a} that there is an offset between the velocity dispersion measurements of ATLAS$^{\rm 3D}$ and SLUGGS data of $\sim20\,{\rm km\,s}^{-1}$. As indicated in the previous section, we have adopted a similar mechanism as employed by \citet{Cappellari15} to compensate for this offset, namely to increase the SLUGGS $\sigma$ values by the offset amount. We here briefly discuss the implications of this.

A suggested cause of the $\sigma$ offset is the binning in the edges of the ATLAS$^{\rm 3D}$ data \citep{Pastorello16, vandeSande17}, resulting in augmented velocity dispersion values. 
If this is the case, then the offset would be greatest in the outer regions of the ATLAS$^{\rm 3D}$ data, coincident with the radial range over which the offset is measured. In this case, increasing the SLUGGS $\sigma$ values may have been unnecessary. Another way in which the offset could be accounted for is to reduce the $\sigma$ values of the ATLAS$^{\rm 3D}$ data to match the inner SLUGGS values. It is then possible, however, that $\sigma$ becomes underestimated in the central region if the offset was caused by binning. 

We run JAM on input data in which the $\sigma$ offset has been implemented both ways: (\textit{i}) increasing the SLUGGS velocity dispersion values by the offset amount, and (\textit{ii}) decreasing the ATLAS$^{\rm 3D}$ velocity dispersion values by the offset amount. We find that this does not have a systematic effect on the $\gamma_{\rm tot}$ retrieved.
As indicated in the final paragraph of Section \ref{sec:PreparationOfData}, the effect of accounting for this offset or not on final $\gamma_{\rm tot}$ values is negligible. 

While the final $\gamma_{\rm tot}$ value is insensitive to the treatment of the velocity dispersion offset, it has a large impact on the retrieved dark matter fractions (when using model 2). The dark matter fractions are systematically larger (in some cases unphysically so, with dark matter fractions in the inner 1 $R_e$ approaching 100 per cent) when increasing the SLUGGS $\sigma$ values to match those of the ATLAS$^{\rm 3D}$ profiles. This is an indication that the outer velocity dispersion values are unphysically high when accounting for the $\sigma$ offset in this manner. Since we are not focusing on dark matter fractions in this paper, we do not address the treatment of this offset to retrieve true measurements. This will need to be addressed in future studies utilising these two datasets to accurately measure dark matter fractions at larger radii using dynamical models.

\subsubsection{NGC 1052 MGE}

In this work, we use \textit{Spitzer} imaging presented by \citet{Forbes17} to derive a stellar MGE for the galaxy NGC 1052, as this galaxy was not part of the ATLAS$^{\rm 3D}$ survey and therefore does not have a literature MGE. The MGE has therefore been derived in a different band (the \textit{Spitzer} $3.6\,\mu{\rm m}$ band as opposed to the $i$ band), and may not be directly comparable with those of the rest of the galaxy sample. This is an important caveat that should be kept in mind when interpreting all results for NGC 1052.
We note, however, that the distribution of mass is expected to be similar in the two bands. We are interested in this distribution and not the absolute scaling of the light. 
We present the MGE for NGC 1052 in Appendix \ref{sec:NGC1052MGE}.  

\subsection{Comparison of method with \citet{Cappellari15}}

Much of the methodology applied in this work is similar to that employed by \citet{Cappellari15} to make measurements of $\gamma_{\rm tot}$ utilising JAM modelling on both the ATLAS$^{\rm 3D}$ and SLUGGS datasets for massive SLUGGS galaxies. There are, however, a few aspects which have been altered for this work, which we clarify here. 

The exact mechanism of dealing with the velocity dispersion offset between the two datasets is slightly different. While we directly adjust SLUGGS $\sigma$ values, \citet{Cappellari15} scaled the $V_{\rm rms}$ values to match those of the ATLAS$^{\rm 3D}$ dataset in overlapping regions. As a result, the velocity dispersion values will be greater in the galaxy outskirts in our models. 

As described in Section \ref{sec:MCMCImplementation}, we combine the two datasets using a hyperparameter method. This contrasts to the \citet{Cappellari15} method, where the two datasets were combined into one kinematic input. To avoid having the spatially dense low-uncertainty ATLAS$^{\rm 3D}$ values overpowering the outer SLUGGS kinematics in the $\chi^2$ calculation, the ATLAS$^{\rm 3D}$ uncertainties were manually increased. The advantage of our method over this is that the true ATLAS$^{\rm 3D}$ uncertainties contribute to the final likelihood calculation. 

We also implement an additional mass model to that of \citet{Cappellari15}. The \citeauthor{Cappellari15} results were derived only using a stellar plus dark matter model (equivalent to our model 2). 
By employing model 1 and model 2, we are able to assess the impact of assuming an inner power law total mass distribution (see Section \ref{sec:PowerLawAssumption}), an assumption employed in gravitational lensing, and in some dynamical modelling studies. 


\section{Modelling}

\subsection{Modelling Results}
\label{sec:ModellingResults}

\begin{figure*}
	\centering
	\includegraphics[trim={0mm, 0mm, 0mm, 0mm}, width=180mm]{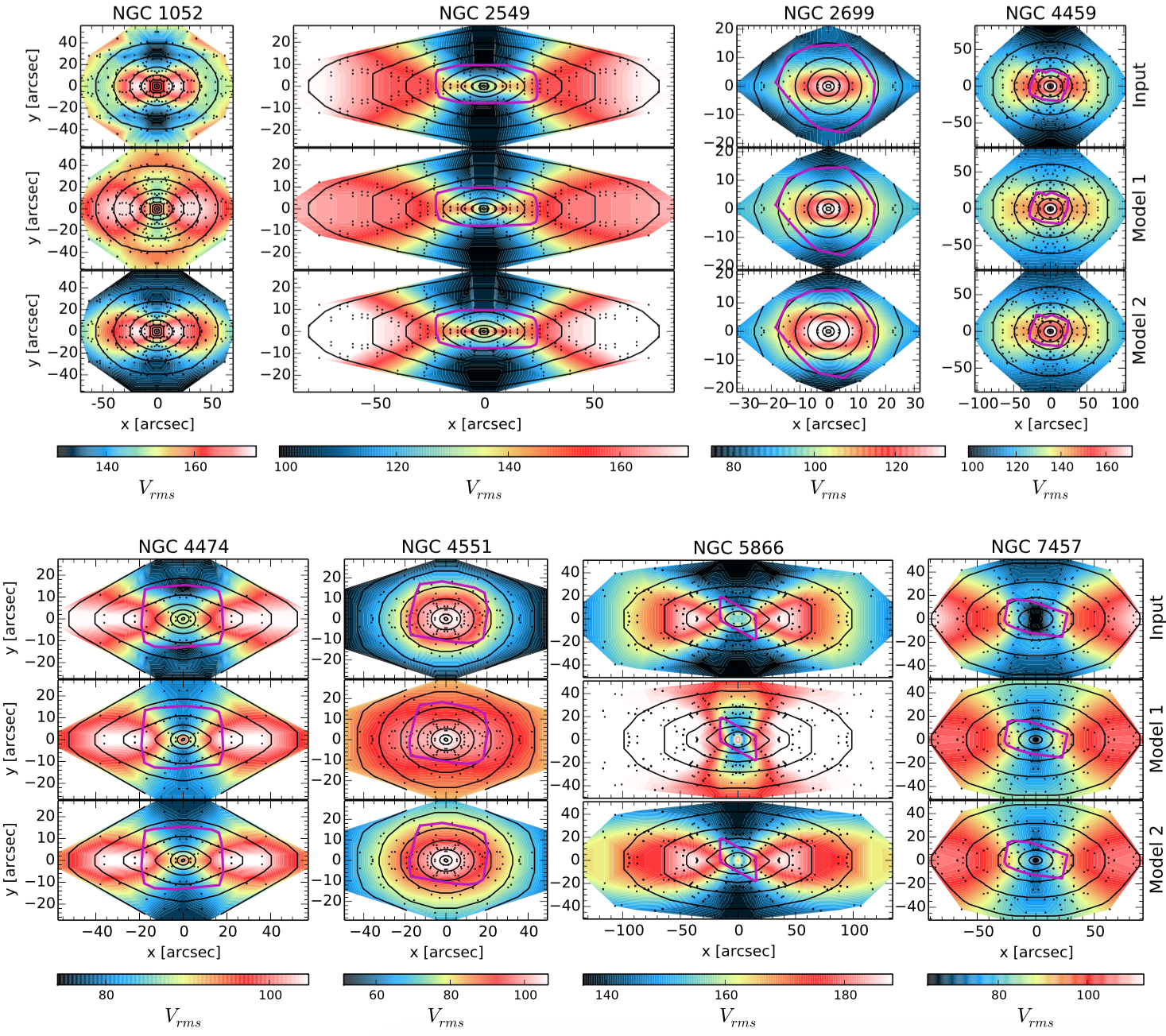}
	\caption{JAM models of NGC 1052, NGC 2549, NGC 2699, NGC 4459, NGC 4474, NGC 4551, NGC 5866 and NGC 7457. The \textit{top} panel for each galaxy presents an interpolated map of the input $V_{\rm rms}$ field, and the other two panels show the output modelled $V_{\rm rms}$ fields. The \textit{middle} panel shows the modelled $V_{\rm rms}$ field according to model 1 (total mass given by a generalised NFW profile), whereas the \textit{bottom} panel shows the modelled field for model 2 (stellar and dark matter mass parametrised separately). Black contours highlight the shape of the galaxy surface brightness in 1 mag intervals, as given by the relevant MGE parametrisation. The colours show the magnitude of the $V_{\rm rms}$ values in km s$^{-1}$, ranging from black as the lowest values, to white as the highest values. The magenta outline in each plot indicates the spatial extent of the ATLAS$^{\rm 3D}$ data used. Small black points indicate the SLUGGS data locations. Note that no ATLAS$^{\rm 3D}$ data were available for NGC 1052. }
	\label{fig:JAM}
\end{figure*}

\begin{table*}
	\centering
	\setlength\extrarowheight{5pt}
	\caption[JAM Parameters1]{Best-fitting parameters using MCMC and JAM for Model 1.  }
	\label{tab:JAMParameters1}
	\begin{threeparttable}
		\begin{tabular}{@{}c|ccccccc}
			\hline
			\hline
			Galaxy  & $i$ & $\beta$  &  log($\rho_{\rm tot}$) &  $\gamma$  &  $\gamma_{\rm tot}$  & $\omega_{\rm SLUGGS}$ & $\omega_{\rm ATLAS3D}$ \\
			& ($^{\circ}$) &  & (1 kpc, $M_{\odot}{\rm pc}^{-3}$) &  &  ($0.1\, - 4\,R_e$)  &  &  \\
			(1) & (2) & (3) & (4) & (5) & (6) & (7) & (8)\\
			\hline
			& \multicolumn{7}{l}{\textit{Galaxies in new sample:}} \\
NGC 1052& $89.81^{+0.14}_{-3.87}$ & $0.10^{+0.02}_{-0.02}$ & $0.092^{+0.004}_{-0.004}$ & $-1.77^{+0.01}_{-0.01}$ & $-1.88^{+0.01}_{-0.01}$ & - & -   \\
NGC 2549& $87.48^{+2.50}_{-1.17}$ & $0.16^{+0.02}_{-0.02}$ & $-0.08^{+0.05}_{-0.06}$ & $-1.97^{+0.01}_{-0.01}$ & $-2.03^{+0.01}_{-0.01}$ & $0.50^{+0.08}_{-0.08}$ & $0.26^{+0.02}_{-0.02}$   \\
NGC 2699& $58.69^{+27.61}_{-1.86}$ & $-0.11^{+0.03}_{-0.03}$ & $-0.29^{+0.05}_{-0.04}$ & $-2.24^{+0.03}_{-0.02}$ & $-2.31^{+0.03}_{-0.02}$ & $0.88^{+0.29}_{-0.21}$ & $0.29^{+0.02}_{-0.02}$   \\
NGC 4459& $54.38^{+2.29}_{-1.99}$ & $0.18^{+0.03}_{-0.03}$ & $0.04^{+0.06}_{-0.07}$ & $-2.07^{+0.01}_{-0.01}$ & $-2.23^{+0.01}_{-0.01}$ & $9.97^{+0.00}_{-0.37}$ & $0.04^{+0.01}_{-0.01}$   \\
NGC 4474& $89.98^{+0.00}_{-1.15}$ & $0.09^{+0.02}_{-0.02}$ & $-0.41^{+0.07}_{-0.05}$ & $-1.98^{+0.02}_{-0.01}$ & $-2.06^{+0.02}_{-0.01}$ & $1.76^{+0.29}_{-0.30}$ & $1.96^{+0.13}_{-0.14}$   \\
NGC 4551& $89.91^{+0.00}_{-5.96}$ & $0.17^{+0.02}_{-0.02}$ & $-0.32^{+0.06}_{-0.06}$ & $-1.94^{+0.02}_{-0.02}$ & $-2.01^{+0.01}_{-0.02}$ & $0.77^{+0.14}_{-0.14}$ & $1.93^{+0.18}_{-0.17}$   \\
NGC 5866& $89.96^{+0.00}_{-0.69}$ & $0.29^{+0.02}_{-0.02}$ & $0.15^{+0.07}_{-0.07}$ & $-1.71^{+0.01}_{-0.02}$ & $-1.85^{+0.01}_{-0.02}$ & $0.23^{+0.10}_{-0.08}$ & $0.36^{+0.03}_{-0.04}$   \\
NGC 7457& $72.0^{+3.38}_{-2.26}$ & $0.20^{+0.05}_{-0.05}$ & $-0.47^{+0.07}_{-0.07}$ & $-1.65^{+0.02}_{-0.02}$ & $-1.80^{+0.02}_{-0.02}$ & $6.96^{+1.18}_{-1.17}$ & $0.27^{+0.03}_{-0.03}$   \\
			\hline
			& \multicolumn{7}{l}{\textit{Other SLUGGS galaxies:}} \\
NGC 821& $86.23^{+3.69}_{-6.14}$ & $0.35^{+0.03}_{-0.03}$ & $0.13^{+0.07}_{-0.07}$ & $-1.94^{+0.02}_{-0.02}$ & $-2.14^{+0.02}_{-0.02}$ & $7.05^{+1.31}_{-1.04}$ & $0.02^{+0.01}_{-0.01}$   \\
NGC 1023& $79.95^{+0.31}_{-0.28}$ & $0.23^{+0.02}_{-0.03}$ & $0.15^{+0.07}_{-0.07}$ & $-2.11^{+0.01}_{-0.01}$ & $-2.24^{+0.01}_{-0.01}$ & $1.27^{+0.11}_{-0.10}$ & $0.003^{+0.011}_{-0.000}$   \\
NGC 2768& $89.96^{+0.00}_{-1.31}$ & $0.33^{+0.02}_{-0.01}$ & $0.29^{+0.05}_{-0.05}$ & $-1.91^{+0.01}_{-0.01}$ & $-2.21^{+0.01}_{-0.01}$ & $8.07^{+0.61}_{-0.61}$ & $0.06^{+0.01}_{-0.01}$   \\
NGC 2974& $64.34^{+3.08}_{-2.00}$ & $-0.01^{+0.09}_{-0.10}$ & $0.27^{+0.13}_{-0.12}$ & $-2.11^{+0.02}_{-0.02}$ & $-2.25^{+0.02}_{-0.01}$ & $0.99^{+0.20}_{-0.17}$ & $0.02^{+0.01}_{-0.01}$   \\
NGC 3115& $89.99^{+0.00}_{-0.37}$ & $-0.07^{+0.01}_{-0.01}$ & $0.225^{+0.004}_{-0.003}$ & $-2.09^{+0.01}_{-0.01}$ & $-2.15^{+0.01}_{-0.01}$ & - & -   \\
NGC 3377& $89.75^{+0.20}_{-2.04}$ & $0.24^{+0.03}_{-0.03}$ & $-0.33^{+0.06}_{-0.06}$ & $-2.07^{+0.02}_{-0.02}$ & $-2.19^{+0.01}_{-0.01}$ & $0.97^{+0.13}_{-0.12}$ & $0.05^{+0.01}_{-0.01}$   \\
NGC 4111& $89.98^{+0.00}_{-0.61}$ & $0.15^{+0.02}_{-0.02}$ & $-0.04^{+0.05}_{-0.05}$ & $-2.1^{+0.01}_{-0.01}$ & $-2.14^{+0.01}_{-0.01}$ & $0.38^{+0.05}_{-0.05}$ & $0.13^{+0.04}_{-0.03}$   \\
NGC 4278& $35.5^{+4.90}_{-3.04}$ & $0.22^{+0.06}_{-0.08}$ & $0.40^{+0.08}_{-0.10}$ & $-2.01^{+0.02}_{-0.02}$ & $-2.14^{+0.02}_{-0.02}$ & $9.97^{+0.00}_{-0.62}$ & $0.03^{+0.02}_{-0.02}$   \\
NGC 4473& $89.94^{+0.00}_{-5.86}$ & $0.28^{+0.03}_{-0.03}$ & $0.23^{+0.07}_{-0.06}$ & $-1.88^{+0.01}_{-0.01}$ & $-2.03^{+0.01}_{-0.01}$ & $4.24^{+0.58}_{-0.53}$ & $0.02^{+0.02}_{-0.02}$   \\
NGC 4494& $89.02^{+0.83}_{-6.95}$ & $0.06^{+0.02}_{-0.01}$ & $0.03^{+0.08}_{-0.07}$ & $-1.87^{+0.01}_{-0.02}$ & $-2.11^{+0.01}_{-0.01}$ & $3.47^{+0.36}_{-0.34}$ & $0.46^{+0.02}_{-0.02}$   \\
NGC 4526& $81.44^{+0.65}_{-0.55}$ & $0.19^{+0.02}_{-0.02}$ & $0.37^{+0.04}_{-0.08}$ & $-1.91^{+0.02}_{-0.01}$ & $-2.08^{+0.01}_{-0.01}$ & $0.47^{+0.07}_{-0.06}$ & $0.06^{+0.01}_{-0.01}$   \\
NGC 4649& $89.87^{+0.00}_{-3.62}$ & $0.04^{+0.02}_{-0.03}$ & $0.52^{+0.08}_{-0.08}$ & $-1.98^{+0.02}_{-0.01}$ & $-2.26^{+0.01}_{-0.01}$ & $5.98^{+0.70}_{-0.74}$ & $0.08^{+0.02}_{-0.02}$   \\
NGC 4697& $89.93^{+0.00}_{-1.35}$ & $0.23^{+0.02}_{-0.02}$ & $0.10^{+0.10}_{-0.10}$ & $-1.97^{+0.01}_{-0.01}$ & $-2.21^{+0.01}_{-0.01}$ & $9.97^{+0.00}_{-0.24}$ & $0.39^{+0.03}_{-0.03}$   \\
			\hline
		\end{tabular}
		\begin{tablenotes}
			\item Columns:
			(1) Galaxy name. 
			(2) Inclination in degrees. 
			(3) Anisotropy. 
			(4) Total mass density at 1 kpc ($M_{\odot}{\rm pc}^{-3}$).
			(5) Inner total mass density slope (free parameter).
			(6) Fitted total mass density slope in $0.1-4\,R_e$ radial range. 
			(7) SLUGGS hyperparameter.
			(8) ATLAS$^{\rm 3D}$ hyperparameter. 
		\end{tablenotes}
	\end{threeparttable}
\end{table*}

\begin{table*}
	\centering
	\setlength\extrarowheight{5pt}
	\begin{threeparttable}	
		\caption[JAM Parameters2]{Best-fitting parameters using MCMC and JAM for Model 2. }
		\label{tab:JAMParameters2}
		
		\begin{tabular}{@{}c | ccccccccc @{}}
			
			\toprule
			\hline
			Galaxy  & $i$ & $\beta$  &  log($\rho_{\rm DM}$) &  $\alpha$ & $\gamma_{\rm tot}$  & $\gamma_{\rm tot}$ & $(M/L)_*$ &  $\omega_{\rm SLUGGS}$ & $\omega_{\rm ATLAS3D}$ \\
			& ($^{\circ}$) &    & ($M_{\odot}{\rm pc}^{-3}$) &  &  ($0.1\, - 4\,R_e$) & ($0.0\, - 4\,R_e$) & &  &    \\
			(1) & (2) & (3) & (4) & (5) & (6) & (7) & (8) & (9) & (10)\\
			\midrule
			&\multicolumn{9}{l}{\textit{Galaxies in new sample:}} \\
NGC 1052& $89.95^{+0.00}_{-5.34}$ & $0.00^{+0.02}_{-0.02}$ & $-0.44^{+0.03}_{-0.08}$ & $-1.50^{+0.00}_{-0.03}$ & $-2.21^{+0.06}_{-0.06}$ & $-1.69^{+0.57}_{-0.57}$ & $1.00^{+0.02}_{-0.00}$ & - & -   \\
NGC 2549& $80.19^{+1.14}_{-0.87}$ & $0.21^{+0.02}_{-0.03}$ & $-0.59^{+0.08}_{-0.09}$ & $-0.80^{+0.13}_{-0.13}$ & $-2.08^{+0.05}_{-0.05}$ & $-1.92^{+0.11}_{-0.11}$ & $4.99^{+0.00}_{-0.09}$ & $0.33^{+0.06}_{-0.05}$ & $0.29^{+0.02}_{-0.02}$   \\
NGC 2699& $37.00^{+3.06}_{-2.17}$ & $0.03^{+0.12}_{-0.10}$ & $-1.56^{+0.68}_{-0.70}$ & $-2.39^{+1.63}_{-0.00}$ & $-2.66^{+0.32}_{-0.32}$ & $-2.11^{+0.87}_{-0.87}$ & $3.82^{+0.15}_{-0.16}$ & $0.97^{+0.24}_{-0.17}$ & $0.30^{+0.02}_{-0.02}$   \\
NGC 4459& $50.06^{+1.38}_{-1.29}$ & $0.18^{+0.03}_{-0.03}$ & $-0.58^{+0.18}_{-0.14}$ & $-1.55^{+0.17}_{-0.22}$ & $-2.22^{+0.22}_{-0.22}$ & $-1.96^{+0.47}_{-0.47}$ & $3.63^{+0.24}_{-0.41}$ & $9.97^{+0.00}_{-0.20}$ & $0.04^{+0.00}_{-0.00}$   \\
NGC 4474& $89.99^{+0.00}_{-1.04}$ & $0.03^{+0.01}_{-0.01}$ & $-0.88^{+0.04}_{-0.04}$ & $-1.52^{+0.09}_{-0.08}$ & $-2.16^{+0.02}_{-0.02}$ & $-1.83^{+0.31}_{-0.31}$ & $2.70^{+0.10}_{-0.11}$ & $1.93^{+0.33}_{-0.29}$ & $2.70^{+0.16}_{-0.15}$   \\
NGC 4551& $86.21^{+3.71}_{-6.23}$ & $0.12^{+0.01}_{-0.01}$ & $-2.99^{+0.49}_{-0.00}$ & $-2.34^{+0.62}_{-0.06}$ & $-2.35^{+0.44}_{-0.44}$ & $-1.85^{+0.94}_{-0.94}$ & $4.72^{+0.05}_{-0.07}$ & $2.53^{+0.40}_{-0.37}$ & $2.42^{+0.17}_{-0.17}$   \\
NGC 5866& $89.96^{+0.00}_{-0.98}$ & $0.21^{+0.01}_{-0.01}$ & $-1.29^{+0.10}_{-0.10}$ & $-0.65^{+0.15}_{-0.15}$ & $-1.94^{+0.42}_{-0.42}$ & $-1.01^{+1.36}_{-1.36}$ & $4.96^{+0.03}_{-0.06}$ & $0.96^{+0.08}_{-0.07}$ & $1.05^{+0.04}_{-0.04}$   \\
NGC 7457& $73.33^{+4.77}_{-2.47}$ & $0.12^{+0.04}_{-0.05}$ & $-0.67^{+0.03}_{-0.03}$ & $-1.44^{+0.04}_{-0.04}$ & $-1.72^{+0.10}_{-0.10}$ & $-1.74^{+0.08}_{-0.08}$ & $1.01^{+0.07}_{-0.0}$ & $7.84^{+1.58}_{-1.47}$ & $0.26^{+0.03}_{-0.03}$   \\
			\hline
			&\multicolumn{9}{l}{\textit{Other SLUGGS galaxies:}} \\
NGC 821& $61.34^{+1.22}_{-0.96}$ & $0.50^{+0.00}_{-0.03}$ & $-0.75^{+0.11}_{-0.09}$ & $-0.68^{+0.09}_{-0.11}$ & $-1.76^{+0.04}_{-0.04}$ & $-1.73^{+0.01}_{-0.01}$ & $4.99^{+0.0}_{-0.1}$ & $2.87^{+0.28}_{-0.27}$ & $0.11^{+0.01}_{-0.01}$   \\
NGC 1023& $79.44^{+0.23}_{-0.26}$ & $0.18^{+0.02}_{-0.02}$ & $-0.14^{+0.06}_{-0.06}$ & $-1.92^{+0.04}_{-0.04}$ & $-2.31^{+0.08}_{-0.08}$ & $-2.04^{+0.34}_{-0.34}$ & $2.32^{+0.13}_{-0.13}$ & $1.57^{+0.11}_{-0.1}$ & $0.0^{+0.01}_{-0.0}$   \\
NGC 2768& $89.95^{+0.0}_{-1.39}$ & $0.26^{+0.02}_{-0.02}$ & $-1.50^{+0.08}_{-0.07}$ & $-0.08^{+0.03}_{-0.09}$ & $-2.03^{+0.04}_{-0.04}$ & $-1.9^{+0.17}_{-0.17}$ & $4.62^{+0.07}_{-0.06}$ & $9.97^{+0.0}_{-0.15}$ & $0.06^{+0.02}_{-0.01}$   \\
NGC 2974& $62.45^{+2.13}_{-1.66}$ & $0.03^{+0.08}_{-0.10}$ & $0.06^{+0.09}_{-0.07}$ & $-2.10^{+0.04}_{-0.04}$ & $-2.30^{+0.10}_{-0.10}$ & $-2.12^{+0.29}_{-0.29}$ & $4.26^{+0.73}_{-1.11}$ & $0.79^{+0.16}_{-0.14}$ & $0.01^{+0.03}_{-0.00}$   \\
NGC 3115& $89.99^{+0.01}_{-2.75}$ & $-0.08^{+0.02}_{-0.02}$ & $0.15^{+0.04}_{-4.12}$ & $-2.05^{+0.03}_{-0.03}$ & $-2.16^{+0.07}_{-0.07}$ & $-2.08^{+0.14}_{-0.14}$ & $1.01^{+3.42}_{-0.00}$ & - & -   \\
NGC 3377& $89.95^{+0.00}_{-2.01}$ & $0.25^{+0.01}_{-0.01}$ & $-1.28^{+0.05}_{-0.03}$ & $-0.08^{+0.03}_{-0.08}$ & $-1.76^{+0.31}_{-0.31}$ & $-1.97^{+0.51}_{-0.51}$ & $2.40^{+0.04}_{-0.04}$ & $9.97^{+0.00}_{-0.53}$ & $0.04^{+0.01}_{-0.01}$   \\
NGC 4111& $89.98^{+0.0}_{-1.31}$ & $0.26^{+0.02}_{-0.02}$ & $-0.3^{+0.06}_{-0.07}$ & $-2.02^{+0.04}_{-0.05}$ & $-1.99^{+0.42}_{-0.42}$ & $-1.90^{+0.51}_{-0.51}$ & $2.45^{+0.21}_{-0.21}$ & $0.47^{+0.05}_{-0.06}$ & $0.15^{+0.03}_{-0.02}$   \\
NGC 4278& $38.79^{+5.71}_{-3.22}$ & $0.15^{+0.05}_{-0.06}$ & $0.35^{+0.00}_{-0.09}$ & $-1.99^{+0.03}_{-0.03}$ & $-2.11^{+0.09}_{-0.09}$ & $-2.02^{+0.17}_{-0.17}$ & $1.01^{+0.25}_{-0.00}$ & $9.97^{+0.00}_{-0.83}$ & $0.03^{+0.01}_{-0.01}$   \\
NGC 4473& $89.94^{+0.00}_{-6.94}$ & $0.26^{+0.02}_{-0.02}$ & $0.13^{+0.00}_{-0.04}$ & $-1.74^{+0.05}_{-0.04}$ & $-2.01^{+0.05}_{-0.05}$ & $-1.83^{+0.23}_{-0.23}$ & $1.01^{+0.06}_{-0.00}$ & $3.84^{+0.45}_{-0.45}$ & $0.02^{+0.01}_{-0.01}$   \\
NGC 4494& $89.87^{+0.13}_{-47.58}$ & $0.04^{+0.01}_{-0.01}$ & $-0.39^{+0.06}_{-0.23}$ & $-1.65^{+0.26}_{-0.06}$ & $-2.16^{+0.20}_{-0.02}$ & $-1.69^{+0.67}_{-0.67}$ & $2.56^{+0.68}_{-0.04}$ & $4.99^{+0.45}_{-0.42}$ & $0.51^{+0.02}_{-0.02}$   \\
NGC 4526& $80.11^{+0.61}_{-0.53}$ & $0.17^{+0.03}_{-0.02}$ & $0.21^{+0.06}_{-0.08}$ & $-1.85^{+0.04}_{-0.05}$ & $-2.08^{+0.15}_{-0.15}$ & $-1.89^{+0.34}_{-0.34}$ & $2.00^{+0.64}_{-0.59}$ & $0.52^{+0.10}_{-0.09}$ & $0.06^{+0.03}_{-0.04}$   \\
NGC 4649& $51.08^{+3.59}_{-2.22}$ & $0.11^{+0.03}_{-0.03}$ & $0.03^{+0.06}_{-0.05}$ & $-1.77^{+0.05}_{-0.06}$ & $-2.32^{+0.12}_{-0.12}$ & $-1.78^{+0.66}_{-0.66}$ & $4.99^{+0.00}_{-0.05}$ & $2.4^{+0.18}_{-0.19}$ & $0.18^{+0.02}_{-0.03}$   \\
NGC 4697& $89.94^{+0.00}_{-1.74}$ & $0.17^{+0.02}_{-0.02}$ & $-1.11^{+0.06}_{-0.07}$ & $-0.75^{+0.06}_{-0.06}$ & $-2.13^{+0.08}_{-0.08}$ & $-1.82^{+0.39}_{-0.39}$ & $4.39^{+0.06}_{-0.05}$ & $9.97^{+0.00}_{-0.48}$ & $0.41^{+0.03}_{-0.03}$   \\
			\hline
			\bottomrule
		\end{tabular}			
		\begin{tablenotes}
			\item Columns:
			(1) Galaxy name.
			(2) Inclination. 
			(3) Anisotropy. 
			(4) Dark matter density at 1 kpc.
			(5) Inner dark matter slope. 
			(6) Fitted slope of the total density profile, measured at $0.1\, - 4\,R_e$. 
			(7) Fitted slope of the total density profile, measured at $0.0\, - 4\,R_e$.
			(8) Stellar mass-to-light ratio (in the $i$ band for all galaxies, except NGC 1052 for which the value is in the \textit{Spitzer} $3.6\,\mu {\rm m}$ band).
			(9) SLUGGS hyperparameter.
			(10) ATLAS$^{\rm 3D}$ hyperparameter.
		\end{tablenotes}
		
	\end{threeparttable}
\end{table*}

We present the final parameters of each of the low stellar mass galaxies (in addition to the other galaxies previously studied by \citealt{Cappellari15}) in Table \ref{tab:JAMParameters1} for model 1, and Table \ref{tab:JAMParameters2} for model 2.  

\begin{figure*}
	\centering
	\includegraphics[trim={0mm, 0mm, 0mm, 0mm}, width=180mm]{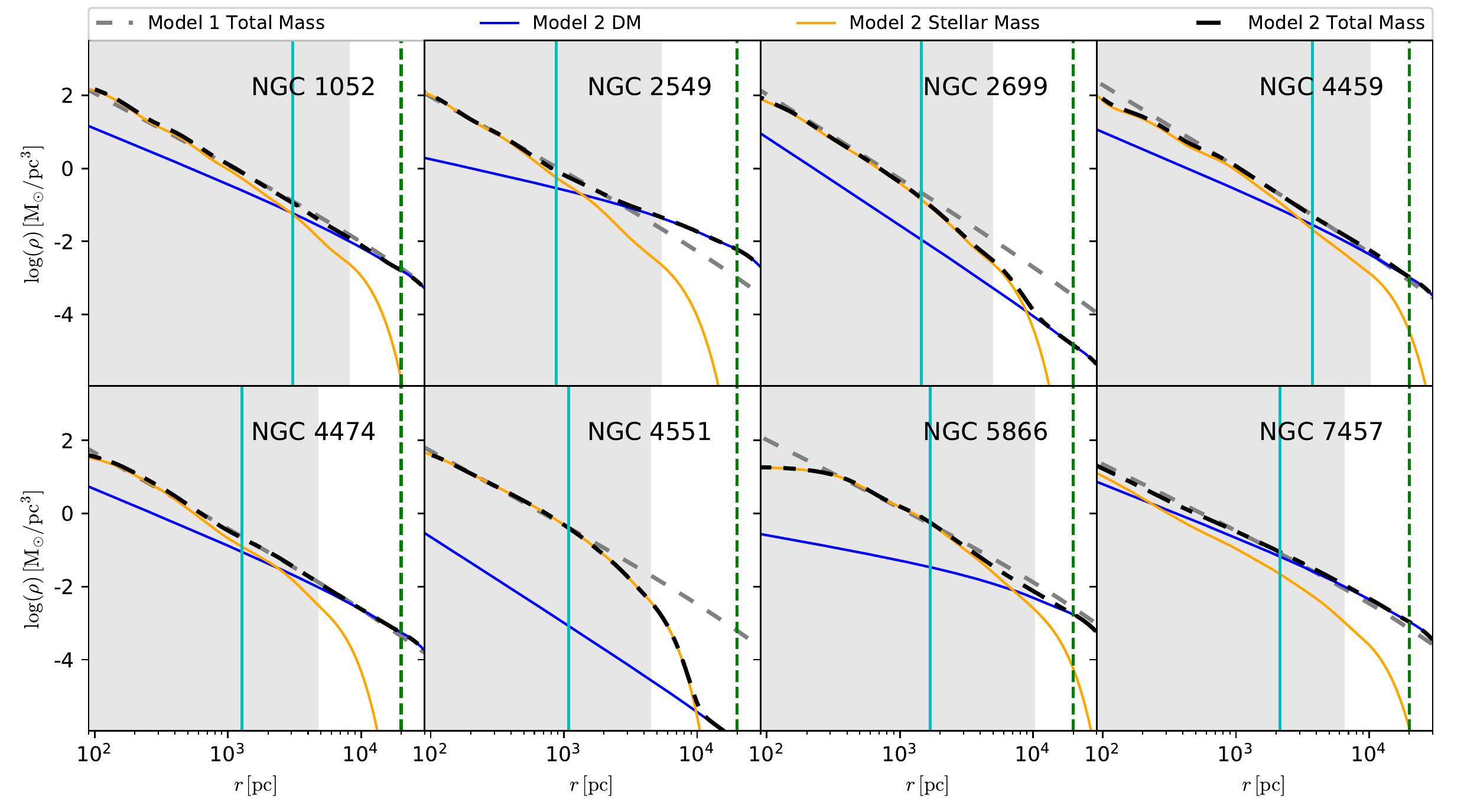}
	\caption{Mass density profiles generated with the separate models. Dashed grey lines show the total-mass density slopes generated by model 1 for each individual galaxy. Since model 2 separates the total mass into stellar and dark components, we plot three lines for model 2. The orange solid line shows the stellar component, while the blue solid line shows the dark matter component. The dashed black line represents the total-mass density slope for model 2. Note how the power law behaviour from model 1 generally follows the behaviour of the total-mass density slope for model 2. The vertical dashed green line shows the assumed break radius (fixed to be 20 kpc) for each galaxy, whereas the vertical solid cyan line indicates the effective radius ($R_e$) for each galaxy. The shaded grey region indicates the radial extent of the observational data used for each galaxy. Note that for galaxies such as NGC 2699 and NGC 4551 that display sudden drops in the model 2 total density profile, these drops occur beyond the radial range for which we have data.  }
	\label{fig:MassProfiles}
\end{figure*}

The input (observed) and model $V_{\rm rms}$ maps for each of the galaxies are shown in Fig. \ref{fig:JAM}. In this figure, for each galaxy, the top panel shows the input kinematics, the middle panel shows the model 1 kinematic output, and the lower panel shown the model 2 kinematic output. The agreement between these $V_{\rm rms}$ maps varies, and is discussed in Section \ref{sec:KinematicFits}. The mass profiles generated by both models for each of the galaxies are plotted in Fig. \ref{fig:MassProfiles}. 

We also apply our modelling techniques to the remaining SLUGGS galaxies, for which measurements using SLUGGS plus ATLAS$^{\rm 3D}$ data have previously been published by \citet{Cappellari15}. We present the JAM models for these galaxies in Appendix \ref{sec:ExtraGalaxiesResults}. 
While the general agreement between our JAM model kinematics and those of \citet{Cappellari15} are good, there are differences present resulting from our implementation as outlined in the previous section. These result not only from the changes in JAM modelling, but also from the manner in which the input $V_{\rm rms}$ field is generated (see Sections \ref{sec:PreparationOfData} and \ref{sec:VelocityDispersionOffsetImpact}). The input $V_{\rm rms}$ maps of NGC 821, NGC 3377, NGC 4111, NGC 4494 and NGC 4649 all display small differences between our study and that of \citet{Cappellari15}. As a consequence, the resulting JAM modelling is also different, highlighting the need for the full sample to be modelled in an entirely consistent manner.

\subsection{Modelling Discusion}
\label{sec:ModellingDiscussion}

\subsubsection{Kinematic fits}
\label{sec:KinematicFits}

The model kinematics produced by JAM are able to recover the kinematic behaviour across the radial extent of the SLUGGS data, in the same manner as was first presented with SLUGGS data by \citet{Cappellari15}. NGC 2549 and NGC 7457 both have $V_{\rm rms}$ lobes that increase in magnitude beyond the radial extent of the data (indicative of the rotational dominance of the galaxy), which is reproduced relatively well by the model $V_{\rm rms}$ maps, although we note that the magnitude of the peak in model 2 for NGC 2549 is higher than for model 1. Conversely, the other galaxies have lobes that peak, and then decline with radius (indicative of the dispersion dominance of the galaxy). The positions of the lobe peaks are generally well recovered in model 2, although the extents are often not an exact match (as is the case for NGC 1052, NGC 4474 and NGC 5866). For NGC 2699 and NGC 4551, the galaxies are so dispersion dominated that only one central $V_{\rm rms}$ peak is observed, rather than two separate lobes.

Both models tend to produce a good kinematic fit to the data for most galaxies\footnote{Out of the \NumberGal$\,$ galaxies analysed in this work, model 2 produces a better kinematic fit for 7 galaxies (in terms of lower residuals), and only produces a worse fit for 3 galaxies. For the remaining galaxies, the fit for the two models is equivalent.  }, but in some cases only model 2 produces a good kinematic fit.
This is visually the most apparent for NGC 1052, NGC 4551 and NGC 5866. 
For NGC 1052, the $V_{\rm rms}$ lobes produced by model 1 extend too far into the outer regions of the galaxy. 
For NGC 4551, model 1 completely fails to produce a good kinematic fit. The predicted $V_{\rm rms}$ value in the outer regions is much too high, whereas the predicted kinematics are a lot better matched for model 2. An explanation for this behaviour can be identified by comparing the resulting total mass density profiles for the two models in Fig. \ref{fig:MassProfiles}. The total mass profile for model 2 is dominated by stellar mass, even in the outer regions, meaning that there is a sharp drop in the density as the stellar mass density declines. This drop in density cannot be replicated by the generalised NFW profile parametrisation of model 1, and hence there is significantly more mass in the outer regions in model 1. This translates directly to greater $V_{\rm rms}$ in the outer regions for model 1. 
The third galaxy for which model 1 fails is NGC 5866. In this case, strong dust structures in the inner region of the galaxy affect both the distribution of stellar light, and the kinematics of the central ATLAS$^{\rm 3D}$ data. The rigid parametrisation of the generalised NFW profile implemented in model 1 is unable to produce a good kinematic fit to the data.

When looking at the kinematic fits for galaxies in the rest of the SLUGGS sample (Fig. \ref{fig:JAM_Extra}), it can be seen that another galaxy with a relatively poor fit is NGC 3377. For this galaxy, the outer region has been better fitted by model 2, whereas the inner region is better fitted by model 1, with neither model able to fit both the inner and outer regions simultaneously. NGC 3377 has an embedded disc, displaying strong downturns in stellar spin and ellipticity in the outer regions \citep[as is clearly shown by its spin--ellipticity ``trail" presented in][]{Bellstedt17b}, which is likely responsible for JAM's inability to return a good kinematic match for the inner and outer regions simultaneously. 

It can be seen that the model kinematics produced by model 2 (separating the stellar and dark matter mass) generally match the observed kinematics better than those of model 1 (in which the total mass distribution is parametrised by a generalised NFW profile). This is particularly clear for galaxies such as NGC 4551 and NGC 5866. 

We highlight that in the radial range of interest, the difference between the total and stellar profiles presented in the model 2 profiles n Fig \ref{fig:MassProfiles} is only very subtle, as has previously been observed by \citet{Romanowsky03}.  

\subsubsection{Consistency of the models}

If the recovered mass distributions for each of the models are ``correct", then we would expect them to be at least consistent with each other. 
We therefore assess whether the two mass models implemented in this work recover the same underlying distribution of mass. 
We do this by comparing the parameters in common between the two models. These parameters are the inclination $i$, the anisotropy $\beta$, and the total-mass density slope $\gamma_{\rm tot}$. 

\begin{figure}
	\centering
	\includegraphics[trim={0mm, 0mm, 0mm, 0mm}, width=90mm]{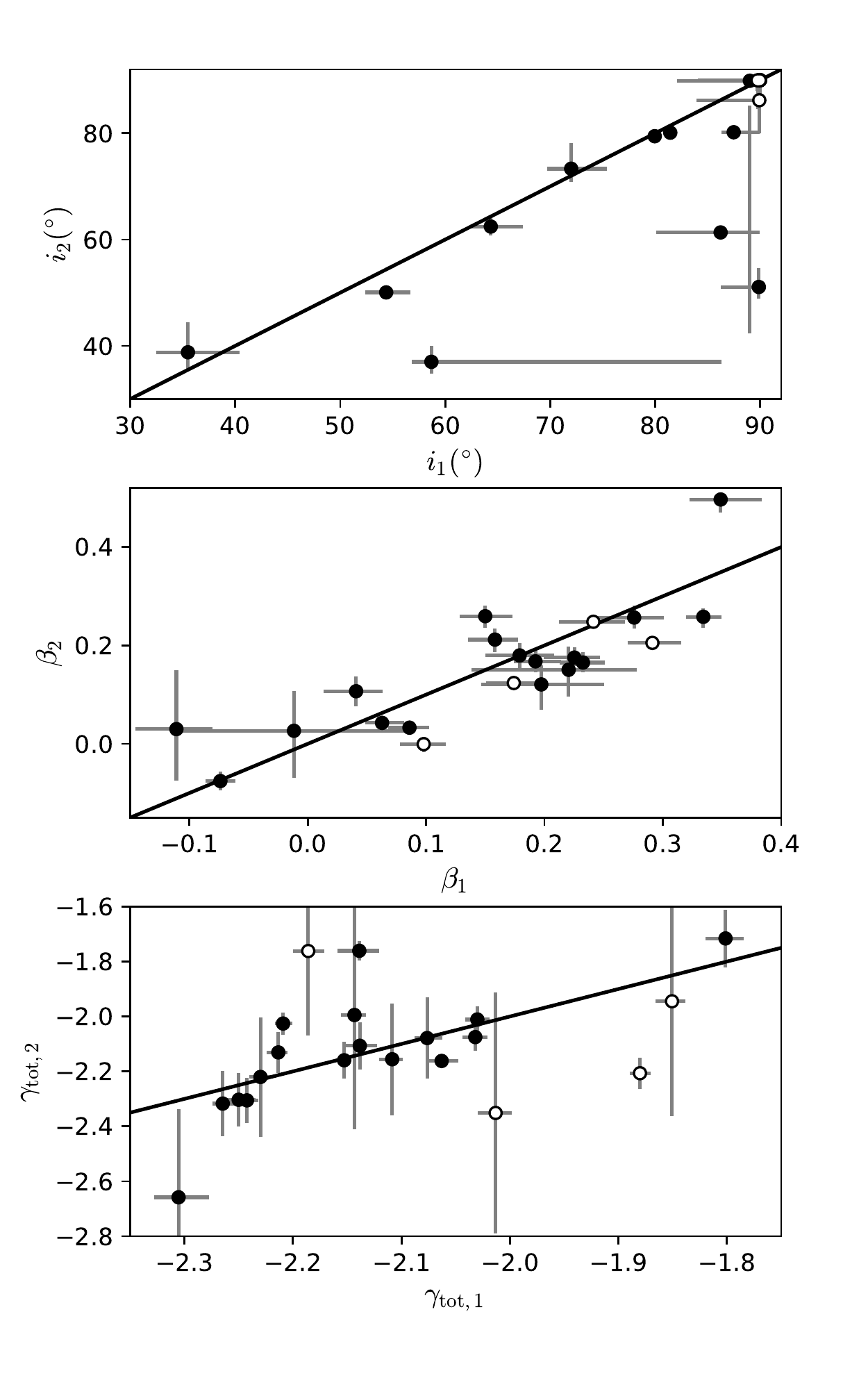}
	\caption{Comparison of common parameters between models 1 and 2. The \textit{top} panel is a comparison of the inclination measurements, the \textit{middle} panel shows the anisotropy measurements, and the \textit{bottom} panel shows the total-mass density profile measurements. Within the bottom panel, $\gamma_{\rm tot}$ values derived from model 2 have been fitted in the radial range $0.1\,R_e-4\,R_e$. Open circles indicate galaxies for which model 1 provides a visually bad fit to the data. }
	\label{fig:JAMModelComparison}
\end{figure}

The comparison between parameters of models 1 and 2 is shown in Fig. \ref{fig:JAMModelComparison}, where the $x$-axes represent model 1 values, and the $y$-axes display model 2 values. 
In galaxies for which one model produces a bad kinematic fit (NGC 1052, NGC 3377, NGC 4551 and NGC 5866), we no longer expect both models to produce the same parameters. We therefore plot these galaxies as open circles in Fig. \ref{fig:JAMModelComparison}. Interestingly, these galaxies are not consistently those with the greatest parameter mismatch. 
In the top panel, a comparison of inclination values shows that model 1 tends to ascribe an edge-on ($90^{\circ}$) inclination to a larger number of galaxies. While there is a general agreement between anisotropy values, it is not exactly one-to-one. Anisotropy is often treated as a `nuisance' parameter \citep[e.g.][]{Cappellari15}, so this lack of agreement is perhaps unsurprising. The rms scatter recovered for $\beta$ is 0.10.  

We plot the fitted $\gamma_{\rm tot}$ values for each model in the bottom panel of Fig. \ref{fig:JAMModelComparison} in which $\gamma_{\rm tot}$ has been fitted in the radial range $0.1-4\,R_e$. For this parameter, the scatter between the two models is 0.19, with $\gamma_{\rm tot}$ values agreeing generally well between models 1 and 2. The fact that the values from the two models generally cluster about the one-to-one line indicates that the same underlying mass distribution is, in fact, recovered by both models, although the uncertainties resulting from the chosen parametrisation are large.

\subsubsection{Selection of final $\gamma_{\rm tot}$ values}
\label{sec:FinalValueSelection}

For the remaining scientific discussion within this paper, we will be referring to the $\gamma_{\rm tot}$ values fitted in the range $0.1\,R_e-R_{\rm max}$\footnote{For galaxies that have $R_{\rm max} > 4R_e$, we fit in the range $0.1-4\,R_e$. While the mean $R_{\rm max}$ value for all galaxies is $4.3R_e$, the range is $2.1-11.3R_e$. } for total mass profiles derived from model 2. This is the same radial range as applied by \citet{Cappellari15}, and does not fit the profile to the innermost region that likely suffers from resolution effects. The minimum fitting radius selected is also the same as the one used by \citet{Poci17}, albeit with a larger maximum radius (given the greater radial extent of our data). 

\section{Measuring total mass density slopes from simulations}
\label{sec:SimulationsGamma}

For each galaxy in the EAGLE samples, we build total mass density profiles in a similar manner to that done for gas profiles in \citet{Stevens17}.  That is, we bin particles in up to 100 spherical shells, where each bin has approximately the same number of particles, with the added conditions that: (\textit{i}) the bin width cannot be less than twice the softening scale of the simulation; and (\textit{ii}) at least 50 particles are in each bin (we have confirmed our results are not sensitive to the precise values used for these conditions).  The radial position of each bin is taken as the mass-weighted mean of the particles in that bin.  Following \citet{Schaller15a}, we exclude the inner parts of profiles that are deemed unreliable,
typically of order a few kpc.
A similar process has been applied to calculate the mass density profiles for the \emph{Magneticum} galaxies, in which the galaxy has been divided into 100 equal-particle bins, in which there are 80 stellar particles per bin. Additionally, the inner-most part of each galaxy is excluded, out to a radius of $2\times r_{\rm softening}$. 

To be consistent with the fitting technique we apply in the observations, we resample the density profiles for the simulated galaxies uniformly in log space. 
The slopes for these resampled density profiles were measured the same way as for the observed galaxies, where the effective radius was taken to be the 2D projected stellar half-mass radius. 
As is apparent in Fig. \ref{fig:ProfilesDifferentModels}, the inner radius of each of the simulated profiles varies as a result of the fixed resolution of the simulations (less massive galaxies have profiles that do not extend as far into the galaxy centre as for the more massive galaxies if normalised to $R_e$). As a result, we cannot fit the profiles in the $0.1-4\,R_e$ radial range for the full sample of simulated galaxies, as was done for the observed galaxies. To make sure that the measurement is consistent across the whole sample, we instead fit the profiles in the radial range $0.4-4\,R_e$, which covers the fittable range for all galaxies in the sample (even the small ones), while avoiding artificial contamination from `cores' resulting from the unresolved central areas of the simulations.

\section{Scientific Context of Results}
\label{sec:ScientificContext}

\subsection{Assumption of a power law slope}
\label{sec:PowerLawAssumption}

It is frequently assumed in studies of mass modelling that the total mass distribution of a galaxy takes on the form of a single power law. In dynamical modelling studies, this assumption is generally relaxed slightly to allow for a double (or `broken') power law (as in our model 1), although this still requires that certain regions of the radial total mass profile are power-law-like. 

We discuss here briefly the validity of the assumption that a total mass profile takes the form of a power law. While the input mass distribution in model 2 does not assume a power law total mass distribution, this assumption is still implicit in the presentation of a $\gamma_{\rm tot}$ value. This discussion is therefore relevant for the results of both models.  

\begin{figure}
	\centering
	\includegraphics[trim={0mm, 0mm, 0mm, 0mm}, width=90mm]{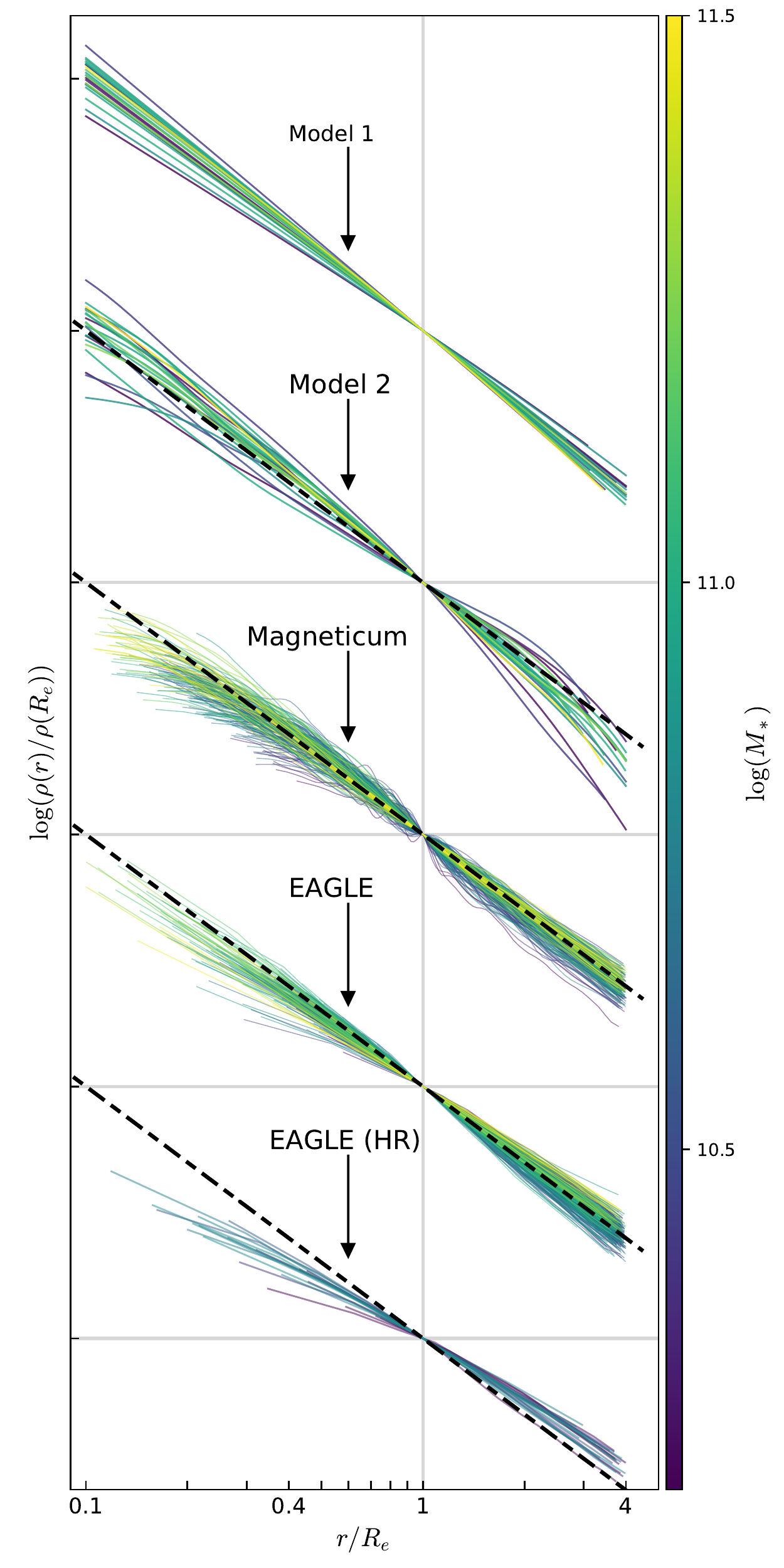}
	\caption{Spherically-averaged mass profiles determined for the SLUGGS galaxies using models 1 and 2, the Magneticum galaxies and the two EAGLE runs (main, and high-resolution). All profiles have been normalised to the value at $1\,R_e$, and vertically offset for the sake of comparison. For each galaxy, the profiles have been plotted in the range $0.1\,-\,4\,R_e$, unless the inner region has been excluded due to resolution effects. We note that the radial range over which we fit $\gamma_{\rm tot}$ for the simulated profiles is $0.4\,-\,4\,R_e$. These profiles clearly show that the observational slopes are steeper than those retrieved by the simulations.  Additionally, the non-power-law-like nature of the high resolution EAGLE runs can be discerned. We plot an isothermal power law ($\gamma_{\rm tot} = -2$) as a dashed black line for each set of profiles for reference.}
	\label{fig:ProfilesDifferentModels}
\end{figure}

\citet{Cappellari15} measured their $\gamma_{\rm tot}$ over the range $0.1-4.0\,R_e$, stating that this is the radial range over which the total-mass density profiles are in the form of a power law. This is also apparent by looking at the model 2 profiles in Fig. \ref{fig:ProfilesDifferentModels}, in which the outer region is generally power law-like, although this behaviour is different in the inner region for some galaxies. 

To compare our chosen mass parametrisations with ``true" total mass density profiles, we compare our profiles with those from simulated galaxies. 
We plot the profiles resulting from the two models for SLUGGS galaxies, in addition to the profiles from the \emph{Magneticum} and EAGLE simulations in Fig. \ref{fig:ProfilesDifferentModels}. 

The \emph{Magneticum} profiles are seen to be essentially isothermal (where the dashed black line for each set of profiles shows the isothermal, $\gamma_{\rm tot} = -2$ power law slope for reference), with power law-like slopes for almost the full radial range of the profile. The exception to this is the very central region of each profile, in which a slight shallowing of the slope can be identified. The range of slopes is similar to those of our model 2, albeit slightly shallower. 

The profiles from the main EAGLE run are slightly shallower than isothermal. Except for some of the lowest-mass galaxies in this run (indicated by the colour of the profiles), which seem to become shallower in the inner regions, most galaxies have slopes that are generally power-law-like. As with the \emph{Magneticum} galaxies, the range in slopes is similar to those of our model 2, albeit even shallower than the \emph{Magneticum} slopes.

\subsection{Total-mass density slopes versus stellar mass}

Fig. \ref{fig:GammaStellarObservations} presents the total-mass density slopes for the SLUGGS galaxies plotted against their stellar mass as orange squares. From the \NumberGal$\,$ galaxies plotted, there is a large scatter with no clear trend to be identified. 
The mean $\gamma_{\rm tot}$ value we recover for SLUGGS galaxies is \FinalGammaValue, consistent with the results from \citet{Cappellari15} and \citet{Serra16}, who similarly utilised data that extend to large radii. 
In their analysis of extended 2D kinematics for a sample of 16 compact elliptical galaxies with a stellar mass range of $9.68 < \log(M_*)/\log({\rm M}_{\odot}) < 11.58$, \citet{Yildirim17} measured a total-mass density slope in the radial range $0.1 - 4R_e$ of $\gamma_{\rm tot} = -2.25$. This is steeper than the measurement we have made for the SLUGGS galaxies, in the same radial range. This is interesting, given that these compact ellipticals are regarded as `relic' galaxies, and hence there is an expectation that a lack of recent merger activity would result in steeper total mass density slopes. 
We compare the results for our galaxies to those from other observational measurements from the literature in Fig. \ref{fig:GammaStellarObservations} in order to expand our understanding of this parameter space. Additionally, we compare these observational measurements to those from simulations in Fig. \ref{fig:GammaStellarSimulations}. 

\subsubsection{Comparison to other observations}
\label{sec:ObservationComparison}

\begin{figure*}
	\centering
	\includegraphics[trim = {0mm, 0mm, 0mm, 0mm}, width=180mm]{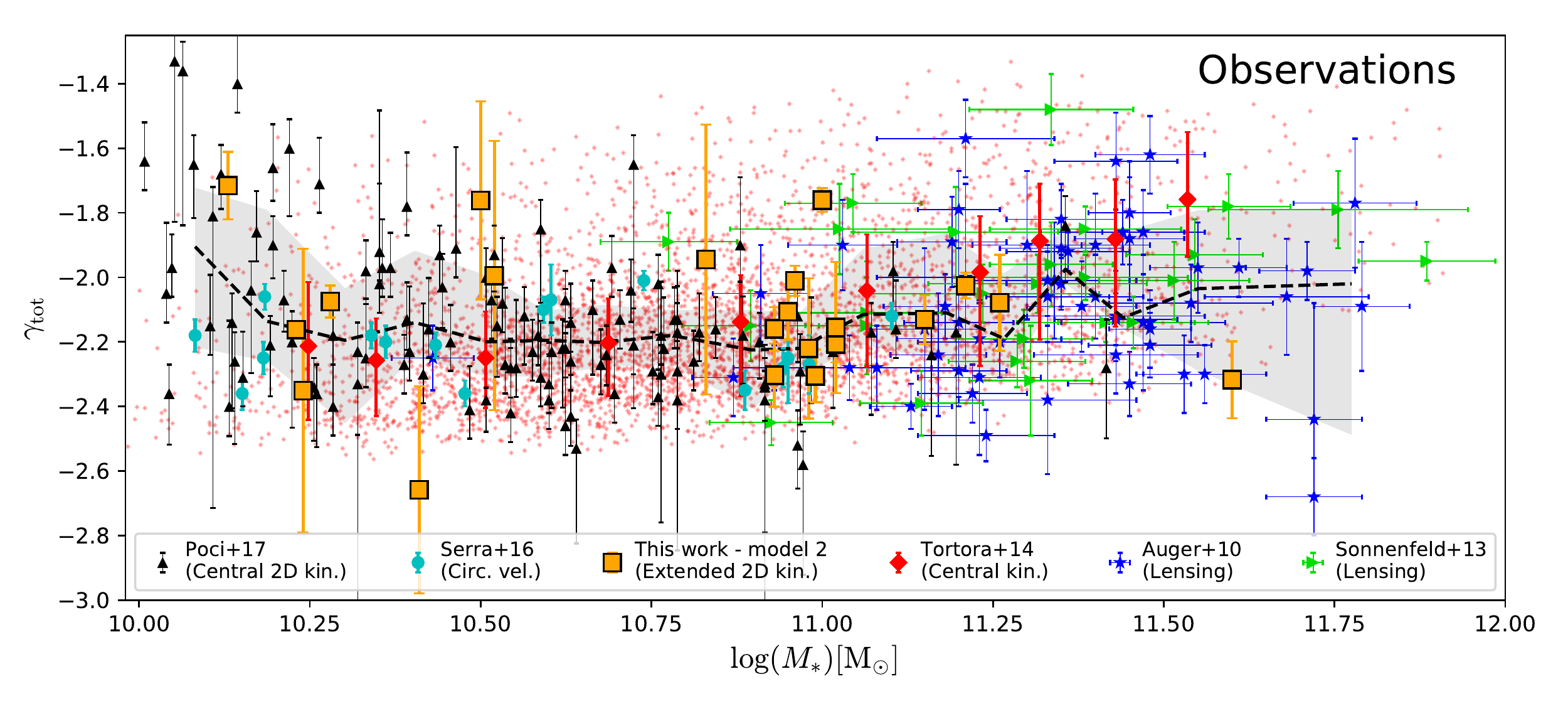}
	\caption{Variation of total-mass density slopes with stellar mass of the SLUGGS galaxies, compared with observations from the literature. The values measured for the SLUGGS galaxies in this work are plotted as orange squares. Observational measurements from \citet{Auger10}, \citet{Sonnenfeld13}, \citet{Tortora14}, \citet{Serra16}, and \citet{Poci17} are included. We note that these studies did not use homogeneous mass models nor radial ranges in calculating total mass density slopes. For the \citet{Tortora14} values, we plot the binned values as large red diamonds, in addition to the full sample, which we plot as the small red points. 
	\citet{Sonnenfeld13} and \citet{Tortora14} stellar mass values have been converted from the original Salpeter values to Chabrier, for consistency. 
	The moving median of all observations is shown by the dashed line, with the 16th--84th percentile range shaded in grey.   }
	\label{fig:GammaStellarObservations}
\end{figure*}

Fig. \ref{fig:GammaStellarObservations} features a comparison between our measurements of $\gamma_{\rm tot}$ and those of other observational studies \citep{Auger10, Sonnenfeld13, Tortora14, Serra16, Poci17}, versus stellar masses. The values we derive are consistent with the results from different studies whose methods for calculating $\gamma_{\rm tot}$ vary, and also the radial extents of the data vary. Below we give a brief description of the techniques implemented by each of these studies, and the differences between them. Broadly, total-mass density slope values are determined through use of either galaxy kinematics or strong gravitational lensing to make mass models. 

Slopes from gravitational lensing studies for massive galaxies are shown in Fig. \ref{fig:GammaStellarObservations} as blue stars \citep{Auger10} and green triangles \citep{Sonnenfeld13}. The two studies targeted similar galaxies, albeit at different redshifts. The galaxies studied by \citet{Auger10} represent `present-day' galaxies up to $z\sim0.4$, whereas \citet{Sonnenfeld13} targeted higher-redshift galaxies, spanning the range $0.1 < z < 0.8$. While we include the \citet{Sonnenfeld13} results for comparison, we do so with the caveat that they are not directly comparable with the other galaxies plotted because of their difference in redshift.   
The published \citet{Auger10} stellar masses were derived using a \citet{Chabrier03} IMF, whereas the \citet{Sonnenfeld13} stellar masses were derived using a \citet{Salpeter55} IMF. To ensure that the plotted stellar masses are consistent, we convert the \citet{Sonnenfeld13} stellar masses to Chabrier masses using the conversion factor of \citet{Madau14}: $M_{*, \rm Chab} = 0.61M_{*, \rm Salp}$. 
The mass distributions within lensing studies are modelled to be spherical, generally probing only the inner $R_e$ of the galaxies. 
\citet{Auger10} noted that accounting for a mild radial anisotropy would lead to measurements of shallower density slopes. We remind the reader that our JAM results return anisotropy values indicative of mild radial anisotropy, but this is not well constrained by our data.

Later studies measuring total-mass density slopes have focused more on dynamics:
\citet{Tortora14} measured the total-mass density slope for galaxies in the SPIDER\footnote{Spheroids Panchromatic Investigation in Different Environmental Regions} \citep{LaBarbera10} and ATLAS$^{\rm 3D}$ surveys, utilising dynamical models produced with the spherical Jeans equations. The modelling was based on a single central velocity dispersion measurement for each galaxy, measured within $R_e/2$.  These results are shown in their binned form as red diamonds in Fig. \ref{fig:GammaStellarObservations}, and results for individual galaxies plotted as small red dots. As for the \citet{Sonnenfeld13} galaxies, we convert the \citet{Tortora14} stellar mass measurements from Salpeter to Chabrier. 

Improving upon dynamical mass models, generated using only a single central kinematic measurement \citet{Serra16} utilised circular velocity ($v_{\rm circ}$) measurements from H\,{\sc i} gas for ATLAS$^{\rm 3D}$ galaxies in the outer regions in addition to inner $v_{\rm circ}$ measurements from JAM \citep{Cappellari13} (measured within 1 $R_e$) to calculate $\gamma_{\rm tot}$. We present these measurements as cyan circles in Fig. \ref{fig:GammaStellarObservations}. Total mass density slopes were derived by probing the circular velocity at two separate radii, since the average total-mass density slope can be determined using the relation $v_{\rm circ} \propto r^{1+\gamma_{\rm tot}/2}$. The H\,{\sc i} measurement was made at varying radii from $4-16\,R_e$, with a median value of $6\,R_e$, hence the radial extent to which the potential has been probed by this method is greater than with the SLUGGS data. 

\citet{Poci17} applied JAM modelling of central 2D kinematics to make $\gamma_{\rm tot}$ measurements for galaxies from the ATLAS$^{\rm 3D}$ survey, which we represent with black triangles. This study therefore used much more detailed kinematic datasets by which to constrain the dynamical mass models. We present their model III results, equivalent to our model 2, in which their slopes are fitted over the radial range $0.1R_e - 1R_e$. For galaxies whose data do not extend to $1\,R_e$, the slopes are fitted to a maximum radius of $R_{\rm max}$. We display the variation between the values from different models as the uncertainty range for each of the \citeauthor{Poci17} galaxies. Since JAM is most effective for galaxies that are viewed edge-on, we choose to plot only those galaxies studied by \citeauthor{Poci17} with an ellipticity $\epsilon > 0.3$ (utilising ellipticity measurements from \citealt{Emsellem11}) as a rough proxy for more highly inclined galaxies. We also exclude any galaxies for which the data quality was noted to be sub-standard by \citet{Cappellari13}, due to the presence of a strong bar, dust, or kinematic twists. 

We show the underlying trends of $\gamma_{\rm tot}$ from all studies with stellar mass in Fig. \ref{fig:GammaStellarObservations} as the median line (black dashed), and the 16th--84th percentiles in the shaded grey region. For the \citet{Tortora14} sample, we use the binned values, rather than the individual galaxies (since this would skew the median line due to the large sample size). We discuss the overall trend with stellar mass in Section \ref{sec:Trends}.

\subsubsection{Comparison to simulations}
\label{sec:SimulationComparison}

\begin{figure*}
	\centering
	\includegraphics[trim = {0mm, 0mm, 0mm, 0mm}, width=180mm]{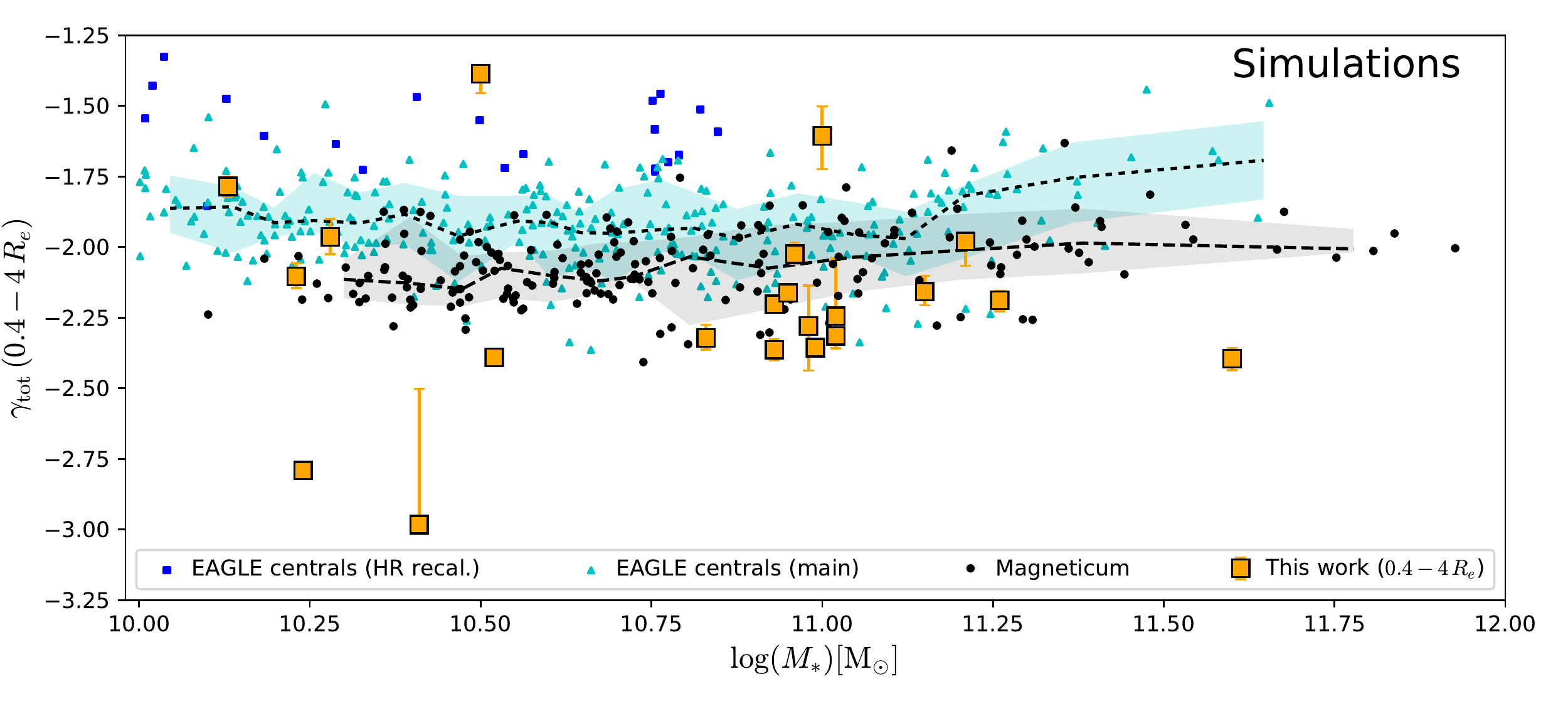}
	\caption{Variation of total-mass density slopes with stellar mass of the SLUGGS galaxies, compared to the simulated values of the \emph{Magneticum} and the EAGLE simulations. Due to inner resolution effects within the simulations, we fit $\gamma_{\rm tot}$ only over the radial range $0.4\,R_e\,-\,4\,R_e$. We therefore plot the observations from our work fitted in the corresponding radial interval. The main EAGLE and \emph{Magneticum} moving medians have been shown as dashed cyan and black lines, with their 16th--84th percentile regions shaded in cyan and grey respectively. We note that at $M_* > 10^{11}\,{\rm M}_{\odot}$, the main EAGLE galaxies display an upturn in $\gamma_{\rm tot}$ values, whereas this is not the case for the \emph{Magneticum} galaxies. We include as blue squares the $\gamma_{\rm tot}$ values for galaxies from the EAGLE high-res run. See Section \ref{sec:EagleHighRes} for a discussion of these galaxies.  }
	\label{fig:GammaStellarSimulations}
\end{figure*}

There are multiple steps that go into deriving a $\gamma_{\rm tot}$ value for both observations and simulations. The total mass profile must first be determined, and only then can $\gamma_{\rm tot}$ be measured from the profile. While mass profiles can be derived directly for simulated galaxy, mass profiles for observational galaxies must be inferred via modelling techniques, as we have shown at length in this paper. The next step of measuring $\gamma_{\rm tot}$ from these derived mass profiles can be done in multiple ways. This is a step which \textit{can} be implemented in the identical way for both observed and simulated mass profiles. 
\citet{Xu17} presented mean $\gamma_{\rm tot}$ values for galaxies in the Illustris simulations, and found the method by which $\gamma_{\rm tot}$ was calculated significantly affects agreement with observations, and that the best agreement was found when fitting a power-law slope to the profiles. We therefore take care to measure the simulated $\gamma_{\rm tot}$ values from mass profiles in the same manner as our observational values, as we have described in Section \ref{sec:SimulationsGamma}. Because the $\gamma_{\rm tot}$ for the simulations was required to be fitted in the radial range $0.4-4\,R_e$, we also now fit the observational $\gamma_{\rm tot}$ values in the same range. This is in addition to the $\gamma_{\rm tot}$ values fitted in the radial range $0.1-4\,R_e$ presented in the previous section. 
We therefore do not expect that the method with which we measure $\gamma_{\rm tot}$ for the simulated galaxies would cause systematic offsets between simulated and observed galaxies.

We compare the galaxies from the \emph{Magneticum} and EAGLE galaxies with our observational measurements in Fig. \ref{fig:GammaStellarSimulations} for the first time in the literature. These simulated galaxies cover a stellar mass range equivalent to our observational data. 
The \emph{Magneticum} values have been plotted as black circles. For the EAGLE simulations, we plot the values for the main run and the recalibrated high-resolution run separately. We separately show the median and percentiles for the \emph{Magneticum} (grey shaded region) and main EAGLE (cyan shaded region) simulated galaxies. The values for the main run are plotted as cyan triangles, and the high-res results are plotted as blue squares. Stellar masses for the \emph{Magneticum} galaxies are measured as the stellar mass within $0.1\,R_{\rm vir}$, whilst stellar masses for the EAGLE simulations are measured as the stellar mass within $R_{\rm 200c}$.

The observed $\gamma_{\rm tot}$ values tend to be $\sim0.1$ steeper when measured in the $0.4-4\,R_e$ radial range, as opposed to the $0.1\,-\,4\,R_e$ radial range implemented for values in Fig. \ref{fig:GammaStellarObservations}. The exceptions to this are NGC 2699, NGC 4111, NGC 4551 and NGC 5866, whose values become significantly steeper in Fig. \ref{fig:GammaStellarSimulations} (by $\sim0.5$). In the case of NGC 4111 and NGC 5866, this is due to the presence of a stellar core. For NGC 2699 and NGC 4551, this steepening is caused by a greater weighting toward the outer region of the profile, in which the stellar profile becomes steeper, and the dark matter profile does not yet compensate for this steepening (likely due to the lower radial extent of our data for these galaxies, see Fig. \ref{fig:MassProfiles}).

The average $\gamma_{\rm tot}$ for observed galaxies is significantly steeper than that of any of the simulations (keeping in mind that the scatter in the observational measurements is larger when measured at the more extended radial range). The offset between the observations and the \emph{Magneticum} values is $\sim 0.1$ (a difference of $1.5\,\sigma$), and the offset between the observations and the main EAGLE run is even greater, at $\sim0.3$ (a difference of $3.4\,\sigma$). 

Given that the slopes have been measured from mass profiles in the same manner, the differences between these samples cannot be explained simply by the region of the galaxy in which we measure $\gamma_{\rm tot}$. Instead, the differences likely originate from one of two other sources: (\textit{i}) The JAM code does not model galaxies in a manner that reflects the true mass profiles of galaxies. (\textit{ii}) The simulations do not produce galaxies that have mass profiles comparable to real galaxies. To identify which of these sources is causing the differences (or even if it is a
combination of both) would require applying JAM modelling to mock observations of simulated galaxies, however such an analysis is beyond the scope of this work\footnote{We note that such a study has previously been done by \cite{Li16}, however, they did not report a comparison of the total mass density slopes. \citet{Remus17} analysed mock observations of simulated galaxies to compare observational and simulated $\gamma_{\rm tot}$ values determined via gravitational lensing techniques.} 

\citet{Xu17} note a similar discrepancy between observations and simulations when comparing the $\gamma_{\rm tot}$ values derived from gravitational lensing to those measured for the Illustris simulation. When measured in the radial range $0.5-2\,R_e$, they found the average $\gamma_{\rm tot}$ for the Illustris galaxies to be $\gamma_{\rm tot} = -2.07\pm0.26$ (compared with the gravitational lensing value of $\gamma_{\rm tot} = -2.078\pm0.027$, \citealt{Auger10}). This is similar to the values we recover for the \emph{Magneticum} simulations (however we highlight that different radial ranges have been used).  

\subsubsection{Trends with stellar mass}
\label{sec:Trends}

In this section we compare the underlying trends that are displayed by the observations in Fig. \ref{fig:GammaStellarObservations}, and the simulations in Fig. \ref{fig:GammaStellarSimulations}.

The median lines of the simulations display different trends with stellar mass. The \emph{Magneticum} galaxies have roughly constant $\gamma_{\rm tot}$ values for varying stellar mass, while the main EAGLE galaxies have constant $\gamma_{\rm tot}$ values only at $M_* < 10^{11.1}\,{\rm M}_{\odot}$. Above this, total-mass slopes become shallower with increasing stellar mass. The relatively flat trend of the \emph{Magneticum} simulations matches relatively well the general flatness of the observational results. 

When assessing the overall trends of the observational measurement, it is important to highlight the caveats that the radial ranges implemented in the fits for $\gamma_{\rm tot}$ are different in each of the studies, and that the samples are a mix of fast- and slow-rotator ETGs. Additionally, multiple studies make use of the ATLAS$^{\rm 3D}$ dataset (including \citealt{Tortora14}, \citealt{Serra16}, \citealt{Poci17} and of course this study), meaning that ATLAS$^{\rm 3D}$ galaxies are over-weighted in the combined results. 
The median line for the observations suggests only slightly shallower total mass density slopes at the massive end, similar to the trend displayed by the \emph{Magneticum} galaxies. At the low-mass end, however, the $\gamma_{\rm tot}$ values also become shallower, with slightly larger scatter. We note that this trend is largely driven by the \citet{Poci17} sample (potentially due to the more relaxed assumptions in the dynamical models applied), since the \citet{Serra16} and \citet{Tortora14} samples individually do not convey this trend.

In summary, total-mass density slopes for the observed galaxies tend to be steeper than isothermal on average, except for the most massive galaxies above $10^{11.2}{\rm M}_{\odot}$, and a scattering of points at the lowest stellar masses below $10^{10.3}{\rm M}_{\odot}$ that have slopes shallower than isothermal. We remind the reader that the shallowing at lower masses is driven by the \citet{Poci17} galaxies. The scatter in results is significantly smaller at $10^{11}{\rm M}_{\odot}$ than at either the high- or low-$M_*$ ends. The simulated galaxies display generally flat trends with stellar mass, except the EAGLE galaxies, which become shallower above $10^{11}{\rm M}_{\odot}$, noting that the simulated galaxies have total-mass density slopes offset from those measured by observations. 

\subsection{EAGLE high-resolution simulations}
\label{sec:EagleHighRes}

In Figs. \ref{fig:ProfilesDifferentModels} and \ref{fig:GammaStellarSimulations} we have included the profiles and total mass density slopes of the high-resolution run of the EAGLE simulations, without yet having discussed the results. We do not regard the profiles from the high-resolution galaxies to be comparable with observed galaxies, for the following reasons.  

The profiles for the high-resolution EAGLE run shown in Fig. \ref{fig:ProfilesDifferentModels} are quite different to those of the main run. The non-power-law nature of these profiles is quite clear when comparing them with the isothermal reference profile, in particular within $1\,R_e$, where the profiles become much shallower. The slopes beyond $1\,R_e$ also show curvature, an indicator that these profiles continue to be non-power-law-like to much greater radii than those of the main run. 


There is a significant offset between the mean $\gamma_{\rm tot}$ values of the two EAGLE runs in Fig. \ref{fig:GammaStellarSimulations}, with the high-resolution run producing shallower slopes by $\sim0.3$.  
It is worth noting that in recalibrating the high-resolution run of EAGLE to reproduce the stellar mass function and size--mass relation, the stellar and AGN feedback strength were both increased.  While one would expect a change in feedback strength to affect the distribution of matter in the centres of galaxies, we have found that both the recalibrated and non-recalibrated high-resolution runs of EAGLE produce consistent central $\gamma_{\rm tot}$ values.  We therefore conclude that feedback parameter values are not the source of this difference.

The star formation law is identical in all runs of EAGLE, despite resolution variations, and has no free parameters. Rather, a fixed value is used for the star formation threshold.  In essence, star formation prescription is a detailed re-description of the Kennicutt--Schmidt law, where star particles are effectively born from gas with hydrogen number densities above $\sim0.1\,{\rm cm}^{-3}$, modulo a metallicity dependence.  In the real Universe, star formation happens at orders of magnitude higher densities than this, in molecular clouds.  Cosmological simulations do not resolve molecular clouds at all though; use of lower density thresholds compensates for this.  In principle, higher-resolution simulations should be able to use higher density thresholds for star formation, but this was not done for EAGLE.  To impose a higher density threshold in a simulation would lead to more centrally concentrated star formation, and hence more centrally concentrated feedback.  This affects the shape of the central potential and thus impacts density profiles.  \citet{Schaller15a} have discussed the lack of dark-matter cores in EAGLE galaxies, citing the importance of the star formation density threshold for this, but noting that other simulations that reproduce cores have not been able to produce the array of observed integrated galaxy properties that EAGLE has. As a point of comparison, the NIHAO simulations impose a density threshold of $\sim10\,{\rm cm}^{-3}$ and recover density cores more in line with observations \citep{Tollet16}.


\subsection{Merger imprint on total mass density slopes}

\begin{figure}
	\centering
	\includegraphics[width=90mm]{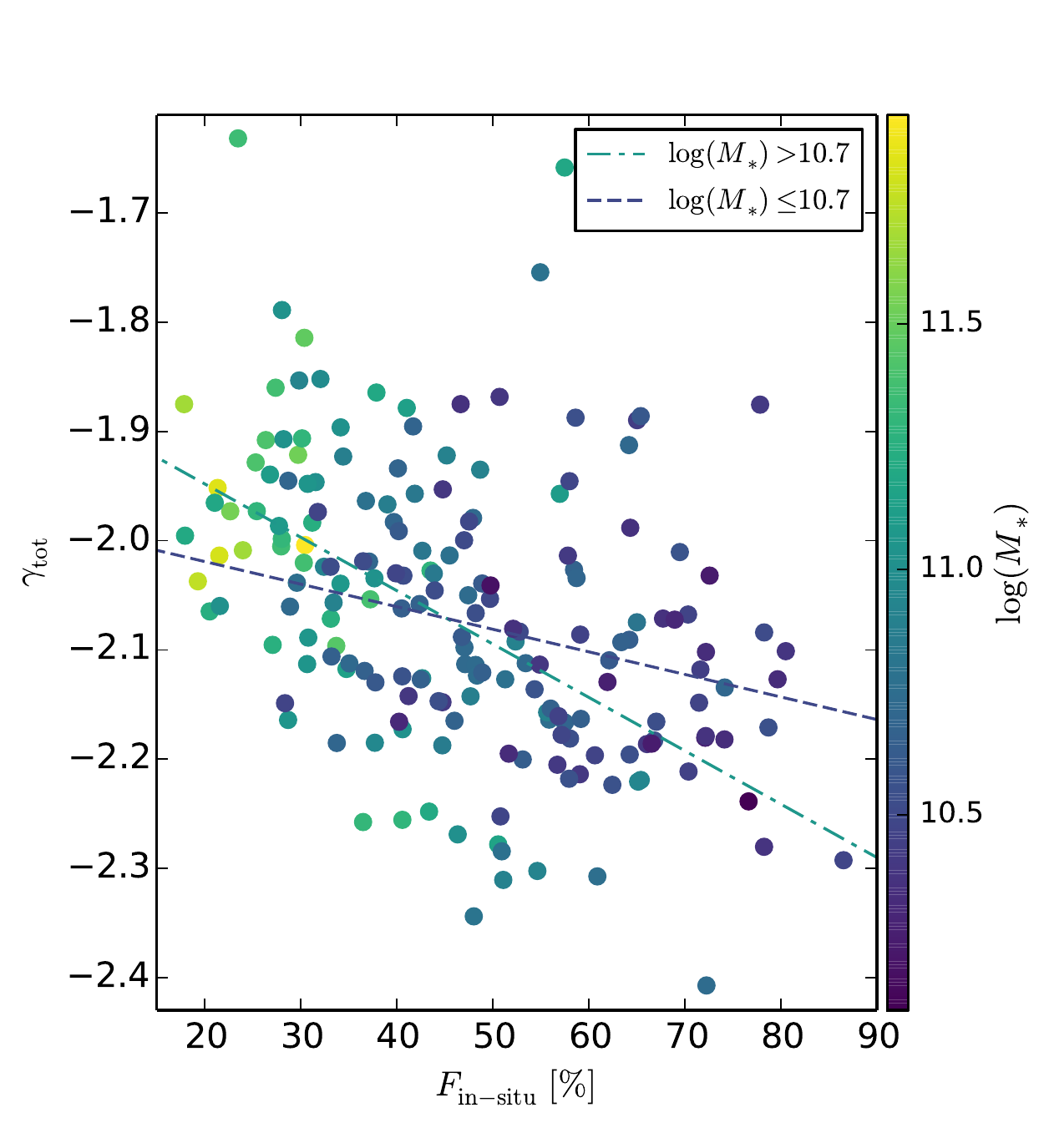}
	\caption{Total mass density profile slopes of \emph{Magneticum} galaxies, plotted against their in-situ fraction. Each point is coloured according to its stellar mass. We split the sample into two stellar mass bins, and fit slopes to these bins. These show that the relation between in-situ fraction of stars and density slope is strongest for the most massive galaxies. The Pearson correlation coefficients for the two stellar mass bins (from most to least massive, respectively), are $-0.44$, and $-0.29$.    }
	\label{fig:GammaInSitu}
\end{figure}

One of the mechanisms one might expect to have a significant influence on the total mass distribution of a galaxy is its merger history. This has previously been investigated by \citet{Remus13}, who used massive galaxies from the \citet{Oser10} simulations to determine that there is a link between the $\gamma_{\rm tot}$ of galaxies and their in-situ stellar fraction (the portion of stars formed within the the main progenitor of a galaxy, as opposed to those formed in merged galaxies), finding that total mass profiles are shallower for more massive galaxies. 

The \citet{Oser10} simulation that was originally analysed to establish the link between $\gamma_{\rm tot}$ and the stellar in-situ fraction did not include the effects of AGN feedback. We investigate whether the link between the stellar in-situ fraction and the total-mass density slope still exists in a cosmological hydrodynamic simulation when this feedback is included. 
When measuring the in-situ fraction for the full sample of \emph{Magneticum} galaxies in Fig. \ref{fig:GammaStellarSimulations}, this trend is much weaker with the presence of AGN feedback, as shown in Fig. \ref{fig:GammaInSitu}. 
This is unsurprising, given that the range in $\gamma_{\rm tot}$ values displayed by the \emph{Magneticum} galaxies is smaller than the range for the \citet{Oser10} galaxies. 
The scatter in this trend is large; for any given value of $\gamma_{\rm tot}$, typical in-situ fractions range between 20 -- 80 percent. This indicates that a measurement of $\gamma_{\rm tot}$ cannot be used as a direct predictor of the in-situ fraction (or conversely, the stellar accreted fraction) of individual galaxies. Fig. \ref{fig:GammaInSitu} suggests that the primary indication of the in-situ fraction is the stellar mass of a galaxy, in that more massive galaxies need many more mergers to grow up to these high masses. Furthermore, as these galaxies usually live in dense environments, their accretion at late times is mostly gas-poor.

The fact that each stellar mass bin for the simulated galaxies displays a trend between in-situ fraction and the total mass density profile slope indicates that the influence of mergers on total-mass distributions is qualitatively ubiquitous. The extent of the merger history, however, is different for galaxies of different masses, with lower-mass galaxies experiencing less dry merger activity than higher-mass galaxies. 
The increased anticorrelation in the $\gamma_{\rm tot}$ -- $F_{\rm in-situ}$ trend highlights that changes in either merger timing, or the ratio of wet/dry mergers affects the mass distributions of galaxies. 
Again, it is interesting to note here that the `relic' galaxies analysed by \citet{Yildirim17}, which are not expected to have experienced recent mergers, display total mass density slopes that are steeper than those that we measure. This would imply that merger activity does have some impact on the measured total mass density slopes. 
Analysing this in detail is beyond the scope of this work.

\subsection{Baryonic--dark matter interplay}

Fig. \ref{fig:GammaStellarObservations} shows a universality in the $\gamma_{\rm tot}$ values measured observationally across a wide stellar mass range ($10^{10} < M_*/{\rm M}_{\odot} < 10^{12}$). Such a universality is not expected when analysing galaxies simulated without AGN feedback, which find that total mass density slopes are shallowest for the most massive galaxies, as is evident from the \citet{Oser10} galaxies presented in \citet{Remus17}. Similarly, simply combining stellar and dark matter density profiles without including mutual interplay would not lead to a $\gamma_{\rm tot}$ independent of stellar mass, since stars and dark matter follow different scaling relations. 
	 
Through use of analytical models, \citet{DelPopolo10} showed that when taking baryonic physics into account, dark matter profile slopes were no longer constant with varying stellar mass of a galaxy, with more massive galaxies displaying steeper dark matter profiles. Without including baryonic physics, these slopes were universal. Given that we observe the \textit{total} mass density profiles to be roughly universal, this indicates that the baryonic and dark matter mass profiles are `conspiring' to maintain a universal $\gamma_{\rm tot}$.

AGN feedback causes the baryonic material within a galaxy to expand and prevents overcooling, resulting in a shallower total mass profile \citep{Remus17}.
There are different physical processes that lead to either expanded or contracted dark matter haloes - as is discussed further by \citet{Dutton15}. Those authors determined that for simulations of massive galaxies, halo expansion was caused by minor merging and stellar mass loss, while halo contraction was caused by dissipational gas accretion (see also \citealt{Remus13} for the effect of gas physics on $\gamma_{\rm tot}$, and \citealt{Sonnenfeld14}). While expansion was caused in simulated galaxies by merger activity, it was noticed that the net effect was still a contraction in the mass distribution, as dissipational activity outweighed the expanding effects of mergers.

The results discussed in this paper highlight that even at opposing ends of the mass scale, in which different shaping mechanisms are expected to dominate, total mass density slopes are roughly constant. This provides an important constraint for simulations, and highlights the need to consider lower-mass galaxies, in addition to massive early-type galaxies when developing models further. Given that we identify differing offsets between mean $\gamma_{\rm tot}$ values of simulations as opposed to observations, the manner in which the baryonic--dark matter interplay is implemented in the simulations has not yet been perfected.

\section{Future Work}
\label{sec:FutureWork}

There are a number of elements in our implementation of JAM that need to be refined in future work, so that constraints on the total mass distribution of galaxies can be improved. 
Owing to the artificial way in which we have dealt with the measured offset in velocity dispersion values between the SLUGGS and ATLAS$^{\rm 3D}$ datasets, we have not been able to produce accurate dark matter fractions or dark matter density slopes, as the outer velocity dispersion values are too high. The origin of such offsets needs to be determined such that they can be correctly accounted for, and to make accurate measurements of the dark matter fractions at different radii. 
Additionally, we have assumed that mass-to-light ratios are constant with radius. In work by \citet{Poci17}, stellar population modelling was done to derive $M/L_*$ profiles for the inner regions of the galaxies, which was used as an input to JAM. Such an approach would improve constraints on the dark matter fractions of galaxies by reducing the degeneracy between the stellar mass-to-light ratio and the dark matter fraction.

There are some galaxies analysed in this work, particularly those with embedded dics such as NGC 3377, for which our JAM models are unable to reproduce the kinematics in both the inner and the outer regions simultaneously. This is possibly because we have assumed constant anisotropy with radius. To improve such models, it would be advantageous to implement a variable anisotropy, particularly for galaxies that display differing kinematic structure in their inner and outer regions.  

It is important to understand whether gravitational lensing and dynamical modelling techniques are probing mass distributions in the same manner. One way in which this can be done is to use techniques such as JAM modelling on massive galaxies for which lensing studies have provided mass distributions. Comparing values derived from each method will yield important insight into the reliability and repeatability of these measurements.  

An analysis by \citet{Janz16} used the total mass density profiles for the 14 galaxies studied by \citet{Cappellari15} to determine that the near-isothermal density profiles were found to be broadly consistent with MOND. With this greater sample of total mass profiles that cover a greater stellar mass range, new tests of MOND and the general mass-discrepancy--acceleration relations can be made. 

In order to make useful comparisons between simulations and observations, it is necessary that exactly the same measurements are made for each. Currently, there is great difficulty in achieving this, since observational measurements are best constrained in the inner regions of galaxies, while the resolution limitations of cosmological simulations mean that measurements for simulated galaxies are best constrained at radii greater than $\sim0.5\,R_e$. This means that while care can be taken to make measurements in the same way, the underlying data are fundamentally different. 
%
\section{Summary and Conclusions}
\label{sec:Conclusion}

We measure total mass density slopes for \NumberGal$\,$ fast rotator ETGs of the SLUGGS survey \citep{Brodie14}, including a greater number of galaxies in the low-mass range than the previous study by \citet{Cappellari15}. These measurements are made by conducting dynamical mass modelling on 2D kinematics that extend to large radii. We apply the JAM mass modelling technique \citep{Cappellari08}, using two different parametrisations of mass distribution to recover the total mass density profiles for individual galaxies. 
We determine that the best fits of the model kinematics to the input kinematics of each galaxy are produced by the model in which the total mass is separated into its stellar and dark matter components, where the dark matter distribution is parametrised as a double power law (as given by Equation \ref{eqn:PowerLaw}). 
We find that the measured total mass density slope values are robust to slight variations within the applied modelling, whereas the separation of stellar and dark matter is much more degenerate, and susceptible to such variations in the modelling. 
In order to be consistent with \citet{Cappellari15}, we measure $\gamma_{\rm tot}$ values for each galaxy by fitting a power law slope to the stellar plus dark matter profile in the radial range $0.1 - 4 R_e$, and find an average total-mass density slope of $\gamma_{\rm tot}=$\FinalGammaValue$\,$ for our sample of \NumberGal$\,$ galaxies, identifying no discernable trend with stellar mass. 

We collate measurements of total mass density slopes from multiple observational studies, for a comprehensive overview of the $\gamma_{\rm tot}$ behaviour across a broad range of stellar masses, utilising an array of datatypes and measurement techniques. Overwhelmingly, the studies all tend to produce consistent results when including the central regions, as shown in Fig. \ref{fig:GammaStellarObservations}. This is despite the fact that the data utilised have varying radial extents for individual galaxies - an indication that these slopes do not display much variation with radius. We find that there tends to be very little trend with stellar mass, and that the slight shallowing of total-mass density slopes we see at lower stellar masses may not be real. All studies generally find total-mass density slopes steeper than isothermal ($\gamma_{\rm tot} < -2$). If the central $0.4\,R_e$ are excluded when measuring the total-mass density slopes, we identify that the measured slopes become steeper, a hint that the inclusion of the central total-mass profile is causing the stability of the observational results. 

What is very clear from our comparison of observational $\gamma_{\rm tot}$ measurements to those of galaxies from the \emph{Magneticum} and EAGLE simulations is that not only do the simulated $\gamma_{\rm tot}$ values display offsets from the observational values to shallower slopes, but the simulations systematically differ between themselves.  
This is interesting, given that the observational results are consistent despite differences in datatypes, measurement techniques, and radial extents probed, and an indication that the processes that affect mass distributions have not yet been captured adequately in cosmological hydrodynamic simulations.  
The total mass density slopes within the \emph{Magneticum} simulations show no significant variation with stellar mass, but in the main EAGLE run, galaxies with stellar masses above $\sim10^{11}{\rm M}_{\odot}$ have slopes that become shallower with increasing stellar mass.

We explore the stellar in-situ fractions of the \emph{Magneticum} galaxies to assess the effect of satellite galaxy accretion on the total mass profiles for galaxies. For the more massive galaxies ($M_* > 10^{10.7} \rm M_{\odot}$), an anticorrelation exists between $\gamma_{\rm tot}$ and $F_{\rm in-situ}$, indicating that galaxies with stellar accreted fractions have shallower slopes. For less massive galaxies, this anticorrelation is also present, albeit weaker. 

The baryonic--dark matter interplay is strongly affected by feedback within galaxies. It is this interplay that results in $\gamma_{\rm tot}$ values being roughly universal across a wide stellar mass range. The ability to match simulated $\gamma_{\rm tot}$ values to observational ones for the whole mass range (including the low-mass) is important to constrain the amount of feedback present in galaxies.


\section{Acknowledgements}

We thank the anonymous referee for their comments and contributions. The data presented herein were obtained at the W. M. Keck Observatory, which is operated as a scientific partnership among the California Institute of Technology, the University of California and the National Aeronautics and Space Administration. The Observatory was made possible by the generous financial support of the W. M. Keck Foundation. The authors wish to recognize and acknowledge the very significant cultural role and reverence that the summit of Maunakea has always had within the indigenous Hawaiian community.  We are most fortunate to have the opportunity to conduct observations from this mountain. 

We acknowledge the Virgo Consortium for making their simulation data available. The EAGLE simulations were performed using the DiRAC-2 facility at Durham, managed by the ICC, and the PRACE facility Curie based in France at TGCC, CEA, Bruyeres-le-Ch\^atel.
The \emph{Magneticum Pathfinder} simulations were partially performed at the Leibniz-Rechenzentrum (project ``pr86re"), and partially supported by the DFG Cluster of Excellence ``Origin and Structure of the Universe". 
The models presented in this paper have relied heavily on use of the g2 supercomputing facility at Swinburne University of Technology. This work made use of the NASA/IPAC Extragalactic Database (NED)\footnote{\url{http://ned.ipac.caltech.edu/}}. 
We thank Crescenzo Tortora for sharing the unbinned data for the full \citet{Tortora14} sample, and we also thank Michele Cappellari for making his code for MGE, LOESS and JAM publicly available\footnote{\url{http://www-astro.physics.ox.ac.uk/\~mxc/software/}}. We acknowledge the use of \texttt{Matplotlib} \citep{Hunter07} for the generation of plots in this paper, and the use of \texttt{ChainConsumer} \citep{Hinton16} to analyse the MCMC outputs. 

SB acknowledges the financial support of the Australian Astronomical Observatory PhD Topup Scholarship. DAF thanks the Australian Research Council for their support via the Discovery Project DP160101608. AJR was supported by National Science Foundation grant AST-1616710 and as a Research Corporation for Science Advancement Cottrell Scholar.   

\bibliographystyle{mnras}
\setlength{\bibsep}{0.0pt}
\bibliography{BibtexMaster}

\appendix

\section{NGC 1052 MGE}
\label{sec:NGC1052MGE}

We present the MGE for NGC 1052 used within this work in Table \ref{tab:1052MGE}.

The surface brightness is converted from counts/pixel$^{-2}$ into magnitude/arcsec$^{2}$ by the following photometric equation:
\begin{equation}
\mu_{3.6\mu{\rm m}} = -2.5\log(C_0) + Z_p + 5\log({\rm PixelScale}),
\end{equation}
where $C_0$ is the surface brightness in counts/pixel$^{-2}$, $Z_P$ is the zero point magnitude (which we take as 17.26), and the pixel scale is 1$\farcs$223 per pixel. 
The conversion from surface brightness into $\rm L_{\odot}\,pc^{-2}$ is then done by:
\begin{equation}
I = \left(\frac{64800}{\pi}\right)^2 10^{0.4 ({\rm M}_{\odot, 3.6\mu{\rm m}} - \mu_{3.6\mu{\rm m}})}
\end{equation}

\begin{table}
	\centering
	\caption[NGC1052MGE]{MGE photometry for NGC 1052 derived from \textit{Spitzer} imaging \citep{Forbes17}. Each row describes the parameters for the individual Gaussians making up the MGE. Columns:
		(1) Central peak. 
		(2) Dispersion.
		(3) Axial ratio. }
	\label{tab:1052MGE}
	\begin{tabular}{ccc}
		\hline
		\hline
		Surface brightness & $\sigma$ & $q$\\
		($\rm L_{\odot}\,pc^{-2}$) & (arcsec) & \\
		(1) & (2) & (3)\\
		\hline
59315.848  &   1.1089  &   0.6070\\
14312.650  &   3.0044  &   0.7934\\
3431.7084  &   6.6863  &   0.8304\\
1434.6350  &   11.8426  &   0.6047\\
953.33742  &   22.0070 &   0.7055\\
55.337784  &   50.9866  &   0.4157\\
105.66361  &   53.3888  &   0.9791\\
		\hline
	\end{tabular}
	
\end{table}

\section{Implementing a free scale radius}
\label{sec:FreeScaleRadius}

Throughout this work, our implementation of the mass parametrisations has fixed the scale radius $r_s$ to 20 kpc in order to reduce the number of free parameters in both models 1 and 2. 

\begin{figure}
	\centering
	\includegraphics[width=90mm]{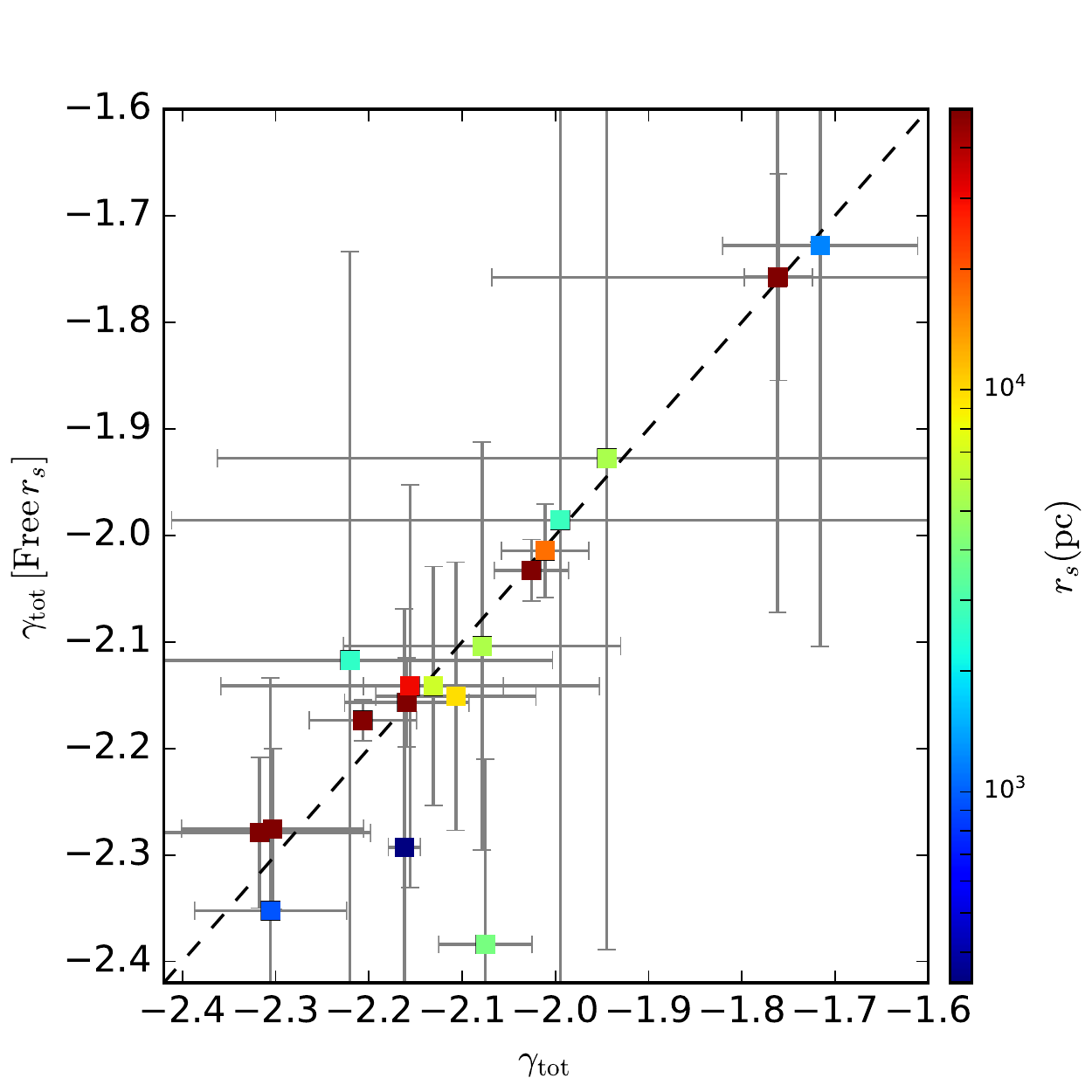}
	\caption{The $\gamma_{\rm tot}$ retrieved when fixing the scale radius, or letting it vary. Each point is coloured according to the $r_s$ value recovered when letting the scale radius be a free parameter. As is expected, the deviation in total mass density slope is greater for smaller $r_s$ values, although the $\gamma_{\rm tot}$ values are still generally consistent within the uncertainties.  }
	\label{fig:FreeScaleRadiusGammaEffect}
\end{figure}

We find that when enabling $r_s$ as a free parameter, there are galaxies for which MCMC is unable to constrain this parameter. Either the value pushes to either end of the prior range, or there is no peak in the posterior distribution at all. For a significant portion of the galaxies, the recovered $r_s$ value when allowing it to be a free parameter is much smaller than 20 kpc. 
The retrieved $\gamma_{\rm tot}$ values for each galaxy are consistent with the values retrieved when the scale radius is fixed to 20 kpc, as is indicated by a comparison of the values in Fig. \ref{fig:FreeScaleRadiusGammaEffect}. 

Despite the fact that the kinematic data used within our study extend to much greater radii than previous studies, the radial extent is not sufficiently large to make accurate constraints on the scale radii. 

\section{Dynamical Models of Other SLUGGS Galaxies}
\label{sec:ExtraGalaxiesResults}

\begin{figure*}
	\centering
	\includegraphics[trim={5mm, 0mm, 5mm, 5mm}, width=175mm]{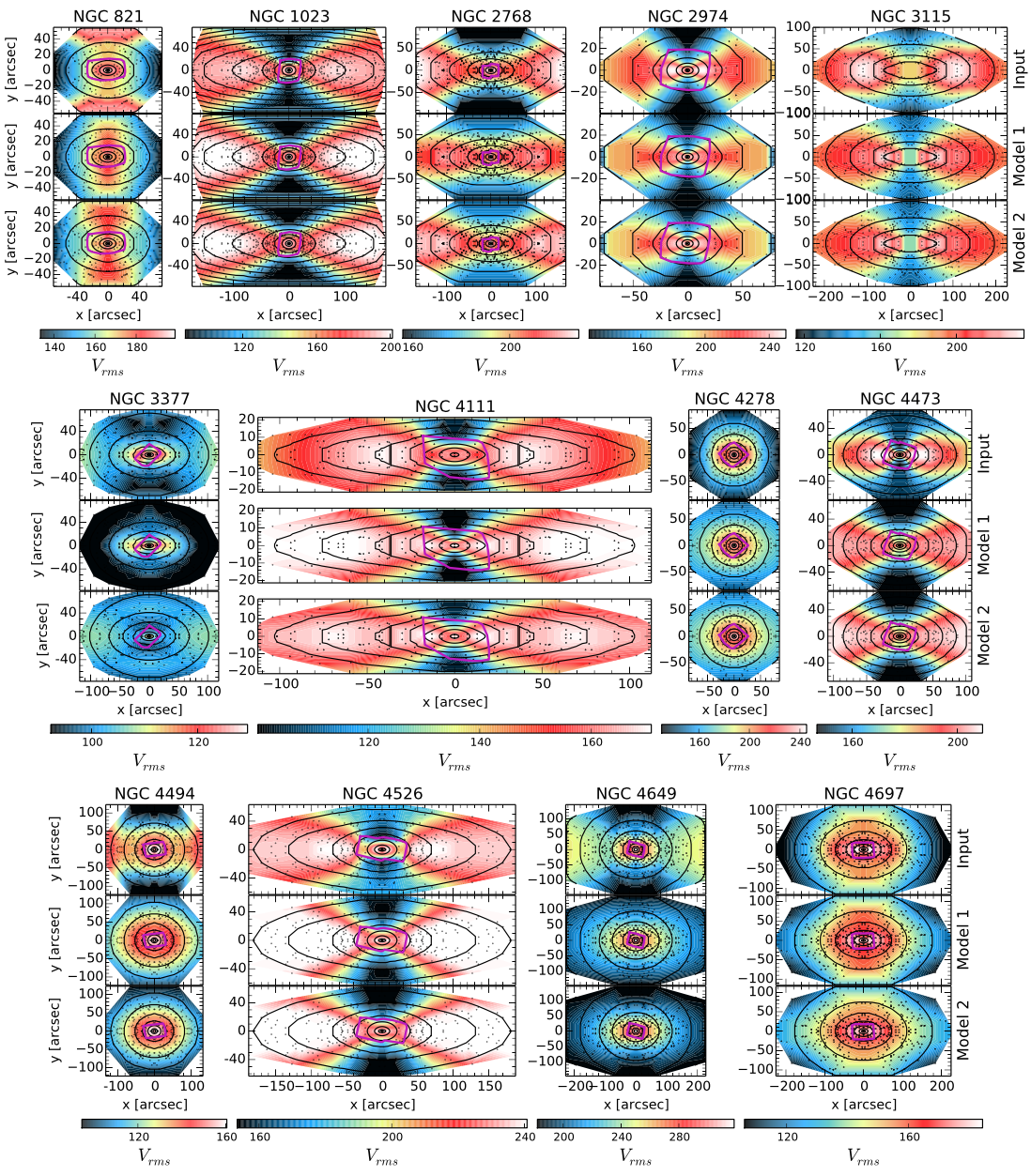}
	\caption{JAM models of galaxies previously studied by \citet{Cappellari15}. The \textit{top} panel for each galaxy presents an interpolated map of the input $V_{\rm rms}$ field, and the other two panels show the output modelled $V_{\rm rms}$ fields. The \textit{middle} panel shows the modelled $V_{\rm rms}$ field according to model 1, whereas the \textit{bottom} panel shows the modelled field for model 2. Black contours highlight the shape of the galaxy surface brightness, as given by the relevant MGE parametrisation. The colours show the magnitude of the $V_{\rm rms}$ values in km s$^{-1}$, ranging from black as the lowest values, and white as the highest values. The magenta line in each plot indicates the spatial extent of the ATLAS$^{\rm 3D}$ data used. Small black points indicate the SLUGGS data locations. Note that no ATLAS$^{\rm 3D}$ data were available for NGC 3115. }
	\label{fig:JAM_Extra}
\end{figure*}

We include in Fig. \ref{fig:JAM_Extra} the JAM kinematic inputs and models for the SLUGGS galaxies previously presented by \citet{Cappellari15}, and that were not presented in Fig. \ref{fig:JAM}. 

\end{document}